\documentclass[twocolumn, linenumber]{aastex62}

\usepackage{amsmath}
\usepackage{array}
\usepackage{natbib}
\usepackage{color}
\usepackage[flushleft]{threeparttable}
\turnoffedit
\newcommand{\mytilde}{\raise.19ex\hbox{$\scriptstyle\sim$}}

\shorttitle{Double Relic Clusters MACS1752 and ZWCL1856}
\shortauthors{Finner et al. }

\begin{document}

\title{Exemplary Merging Clusters: Weak-lensing and X-ray Analysis of the Double Radio Relic, Merging Galaxy Clusters MACS J1752.0+4440 and ZWCL 1856.8+6616}

\correspondingauthor{M. James Jee}
\email{mkjee@yonsei.ac.kr}
\email{kylefinner@gmail.com}
\author{Kyle Finner}
\affil{Department of Astronomy, Yonsei University, 50 Yonsei-ro, Seoul 03722, Korea}

\author{Kim HyeongHan}
\affil{Department of Astronomy, Yonsei University, 50 Yonsei-ro, Seoul 03722, Korea}

\author{M. James Jee}
\affil{Department of Astronomy, Yonsei University, 50 Yonsei-ro, Seoul 03722, Korea}
\affil{Department of Physics, University of California, Davis, One Shields Avenue, Davis, CA 95616, USA}

\author{David Wittman}
\affil{Department of Physics, University of California, Davis, One Shields Avenue, Davis, CA 95616, USA}

\author{William R. Forman}
\affil{Smithsonian Astrophysical Observatory, Harvard-Smithsonian Center for Astrophysics, 60 Garden St., Cambridge, MA 02138, USA}

\author{Reinout J. van Weeren}
\affil{Leiden Observatory, Leiden University, PO Box 9513, 2300
RA Leiden, The Netherlands}

\author{Nathan R. Golovich}
\affil{Lawrence Livermore National Laboratory, 7000 East Ave., Livermore, CA 94550, USA} 

\author{William A. Dawson}
\affil{Lawrence Livermore National Laboratory, 7000 East Ave., Livermore, CA 94550, USA}

\author{Alexander Jones}
\affil{Hamburger Sternwarte, University of Hamburg, Gojenbergsweg 112, Hamburg 21029, Germany}

\author{Francesco de Gasperin}
\affil{Hamburger Sternwarte, University of Hamburg, Gojenbergsweg 112, Hamburg 21029, Germany}

\author{Christine Jones}
\affil{Smithsonian Astrophysical Observatory, Harvard-Smithsonian Center for Astrophysics, 60 Garden St., Cambridge, MA 02138, USA}

\begin{abstract}
The investigation of merging galaxy clusters that exhibit radio relics is strengthening our understanding of the formation and evolution of galaxy clusters, the nature of dark matter, the intracluster medium, and astrophysical particle acceleration. Each merging cluster provides only a single view of the cluster formation process and the variety of merging clusters is vast. Clusters hosting double radio relics are rare and extremely important because they allow tight constraints on the merger scenario. We present a weak-lensing and X-ray analysis of MACSJ1752.0+4440 ($z$=0.365) and ZWCL1856.8+6616 ($z$=0.304), two double radio relic clusters. Our weak-lensing mass estimates show that each cluster is a major merger with approximately 1:1 mass ratio. The total mass of MACSJ1752.0+4440 (ZWCL1856.8+6616) is $M_{200}=14.7^{+3.8}_{-3.3}\times10^{14}\ $M$_\odot$ ($M_{200}=2.4^{+0.9}_{-0.7}\times10^{14}\ $M$_\odot$). We find that these two clusters have comparable features in their weak-lensing and gas distributions, even though the systems have vastly different total masses. From the likeness of the X-ray morphologies and the remarkable symmetry of the radio relics, we propose that both systems underwent nearly head-on collisions. However, revelations from the hot-gas features and our multiwavelength data analysis suggest that ZWCL1856.8+6618 is likely at a later merger phase than MACSJ1752.0+4440. We postulate that the SW radio relic in MACSJ1752.0+4440 is a result of particle re-acceleration.
\end{abstract}


\renewcommand{\arraystretch}{1.5}
\section{Introduction}
Galaxy clusters gain mass primarily by mergers with other galaxy clusters and groups. The most energetic of these events occurs when two galaxy clusters undergo a major merger. Major cluster mergers are ideal laboratories to study high energy astrophysics such as the nature of dark matter and particle acceleration. However, because galaxy clusters are at the high-mass end of the hierarchical formation of the large scale structure, major cluster mergers are rare.

An indication of a merging event can be detected in the intracluster medium (ICM). Emitting thermal bremsstrahlung radiation at X-ray wavelengths, the ICM of merging clusters tends to show a disturbed morphology. The disruption is caused by the hydrodynamical properties of the ICM and is observed in some famous merging clusters such as the Bullet cluster \citep{2004markevitch}, the Toothbrush cluster \citep{2013ogrean_tooth}, and the Sausage cluster \citep{2013ogrean_sausage}. Further investigation with state-of-the-art X-ray telescopes can reveal finer details that trace past merging events such as stripped gas, cold fronts, sloshing, and shocks \citep{2007markevitch, 2010ghizzardi}. In rare cases, where the ram-pressure force on one ICM from the other exceeds the gravitational force, the ICM can dissociate from the dark matter \citep[e.g.][]{2008bradac, 2012dawson, 2012jee, 2016randall}. The projected offset is maximized in clusters that are merging in the plane of the sky.

A primary indicator of a cluster merger is the detection of a shock. Shocks form when the collision velocity of the merging clusters exceeds the sound speed of the ICM. Investigating \mytilde2:1 mass-ratio mergers in cosmological simulations, \cite{2018ha} specified that two types of shocks can form during a cluster merger. As the ICM from each cluster begins to interact, equatorial shocks form and propagate perpendicular to the merger axis \citep[e.g.][]{2019gu}. Near pericenter passage, axial shocks form and propagate ahead of the clusters along the merger axis. Axial shocks are also referred to as bow shocks \citep{1998ricker, 2002sarazin, 2002markevitch} because of their location ahead of the cluster. The geometry of these shocks can be used to constrain the collision axis of the merger.

Merger shocks are observed in X-ray as a discontinuity in the ICM surface brightness or a jump in temperature from the pre- to the post-shock region, \edit2{with the latter being clear evidence}. However, the detection of shocks in X-ray wavelengths is hampered by the low brightness of the ICM outside the cluster core. Fortunately, shocks can also be detected at radio frequencies. Radio relics are produced by synchrotron emission from charged particles that are accelerated in merger-induced shocks \citep[for reviews of radio relics see][]{2012feretti, 2019vanweeren} by diffusive shock acceleration \citep[DSA;][]{1978bell, 1983drury, 2001malkov}. Radio relics tend to be megaparsec in size and are often found on the periphery of merging clusters. To date, the number of clusters observed that host radio relics is about 70, which makes them much rarer than merging clusters \citep[for a list see][]{2019vanweeren}. One reason for the rarity is because axial shocks that face the observer have low surface brightness \citep{2013skillman}. Therefore, radio relics are preferentially found in mergers that are occurring near the plane of the sky \citep[][]{2019bgolovich}. Another contributor to their rarity may be the poor acceleration efficiency of DSA in weak shocks \citep{2005kang, 2020botteon}. To overcome the poor acceleration efficiency of DSA, the re-acceleration model \citep{2011kang, 2012kang, 2013pinzke}, where a population of non-thermal seed electrons is re-accelerated by the shock, may be required. Evidence for particle re-acceleration has been compounding recently with clusters such as PLCKG287.0+32.9 \citep{2014bonafede} and Abell 3411 \citep{2017vanweeren}. In the latter study, a connection of radio emission between a cluster radio galaxy and the radio relic was found.

The rarest of the relic phenomena are double relics that are equidistant from the cluster center. As explained in \cite{2019vanweeren}, double radio relics are defined as shocks formed from a single merger event. Having two probes of the same merging event is an advantage that can be used to put tight constraints on the merger scenario. To date, about 12 double radio relic clusters have been identified. The two clusters that are presented in this work are double radio relic clusters.  

The Merging Cluster Collaboration\footnote{http://www.mergingclustercollaboration.org/} is reconstructing the collisions of merging galaxy clusters that exhibit radio relics to investigate the physics of galaxy clusters. An overview of the merging cluster sample can be found in \cite{2019agolovich}. We are utilizing multiwavelength data to probe the dark matter, ICM, and galaxies so that the components of the merger can be meticulously modeled. We rely on weak lensing (WL) to characterize the mass distributions and estimate subcluster masses in these disturbed systems. The masses that we derive from WL are crucial inputs to our merger simulations \citep[][]{2020wonki}. In this paper, we present the first WL analysis of the double radio relic clusters MACSJ1752.0+4440 (MACS1752) and ZWCL1856.8+6616 (ZWCL1856). 

MACS1752 ($z$=0.365) was discovered in the Massive Cluster Survey \citep{2001ebeling}. \cite{2003edge} found it to be a radio-bright cluster through a joint analysis of the Westerbork (WSRT) Northern Sky Survey \citep{1997rengelink} and ROSAT All-Sky Survey Bright Source Catalog \citep{1999voges}. Follow-up WSRT observations in L and S band \citep{2012vanweeren} revealed that MACS1752 is a merging galaxy cluster with double, megaparsec-sized radio relics and a radio halo. The radio study measured the largest linear size (LLS) of the NE relic to be 1.35 Mpc and the SW relic to be 0.86 Mpc. Integrated spectral indices of the relics were reported as $-1.16^{+0.03}_{-0.03}$ in the NE and $-1.10^{+0.05}_{-0.05}$ in the SW. \cite{2012bonafede} confirmed the presence of the two radio relics and radio halo in Giant Metrewave Radio Telescope (GMRT) observations. Setting the midpoint between the X-ray brightness peaks as the cluster center, they estimated the projected distance to the relics to be 1.13 Mpc and 0.91 Mpc to the NE and SW, respectively. Joining the GMRT and WSRT observations, \cite{2012bonafede} measured consistent integrated spectral indices and reported injection spectral indices of \mytilde0.6 and 0.8 and Mach numbers of 4.6 and 2.8 for the NE and SW relics, respectively. Both relics were shown to have polarization peaking at \mytilde 40\%. They found a spectral index of $-1.33\pm0.07$ for the radio halo. They note that the NE relic shows a clear \edit2{cluster-centric} spectral steepening, a telltale sign of a merger origin of the radio emission, that resembles the steepening found in CIZAJ2242.8+5301 (the Sausage cluster). The Sunyaev-Zel'dovich effect (SZE) mass of the cluster is estimated to be $M_{500} = 6.7^{+0.4}_{-0.5} \times 10^{14}\ $M$_\odot$ from \cite{2016planck}. MACS1752 has also been the subject of numerical simulations in order to investigate the efficiency of electron and proton acceleration in merger shocks \citep{2015vazza} as well as to constrain the efficiency of cosmic-ray acceleration \citep{2016vazza}. In the latter study, a cosmological zoom-in simulation is presented that reproduces the X-ray and radio morphology.

ZWCL1856 ($z$=0.304) \citep{1961zwicky} (PSZ1 G096.89+24.17) was detected in the Planck SZ Survey and reported to have a mass $M_{500}=4.7^{+0.3}_{-0.3}\times10^{14}\ \text{M}_\odot$ \citep{2016planck}. Cross matching WSRT observations with galaxy cluster catalogs, \cite{2014degasperin} discovered that ZWCL1856 contains two radio relics. They estimated the northern relic LLS to be 0.9 Mpc in extent and the southern to be 1.4 Mpc. Taking the brightness peak of the smoothed ROSAT X-ray image as the center of the cluster, they found the distance to the northern relic to be 0.77 Mpc and the southern relic to be 1.15 Mpc. The spectral index of the radio emission was not reported because of a lack of radio data. 

The symmetry of these two clusters provides the ideal configuration to study the connection between cluster mergers and radio relics. The aim of this study is to enhance our understanding of the merger scenarios that created these double radio relics using multiwavelength observations and to provide constraints that can be used in future simulations. Utilizing deep Subaru and \textit{HST} observations of the galaxy clusters, we perform a precision WL analysis of the cluster mass and combine the WL results with X-ray and radio observations to seek a deeper understanding of the merging process.

We outline the observations and data reduction in Section \ref{sec:data_reduction}. Weak-lensing theory, shape measurement, source selection, and source redshift estimation are in Section \ref{sec:wl_method}. Section \ref{sec:results} presents WL mass reconstruction, mass estimations, and X-ray analysis. We discuss the mass of each substructure and provide evidence for updated merger scenarios in Section \ref{sec:discussion}.

Cluster masses are defined as the mass within $R_{200}$, where $R_{200}$ is the radius of a sphere within which the average density is 200 times the critical density of the universe at the cluster redshift. We assume a flat $\Lambda$CDM cosmology with $H_0=70$ km s$^{-1}$ Mpc$^{-1}$, $\Omega_\Lambda=0.7$, and $\Omega_m=0.3$. For the stated cosmology, the scales are 5.123 kpc arcsec$^{-1}$ and 4.535 kpc arcsec$^{-1}$ for $z$=0.365 (MACS1752) and $z$=0.304 (ZWCL1856), respectively.  

\begin{figure*}[ht!]
    \centering
    \includegraphics[width=\textwidth]{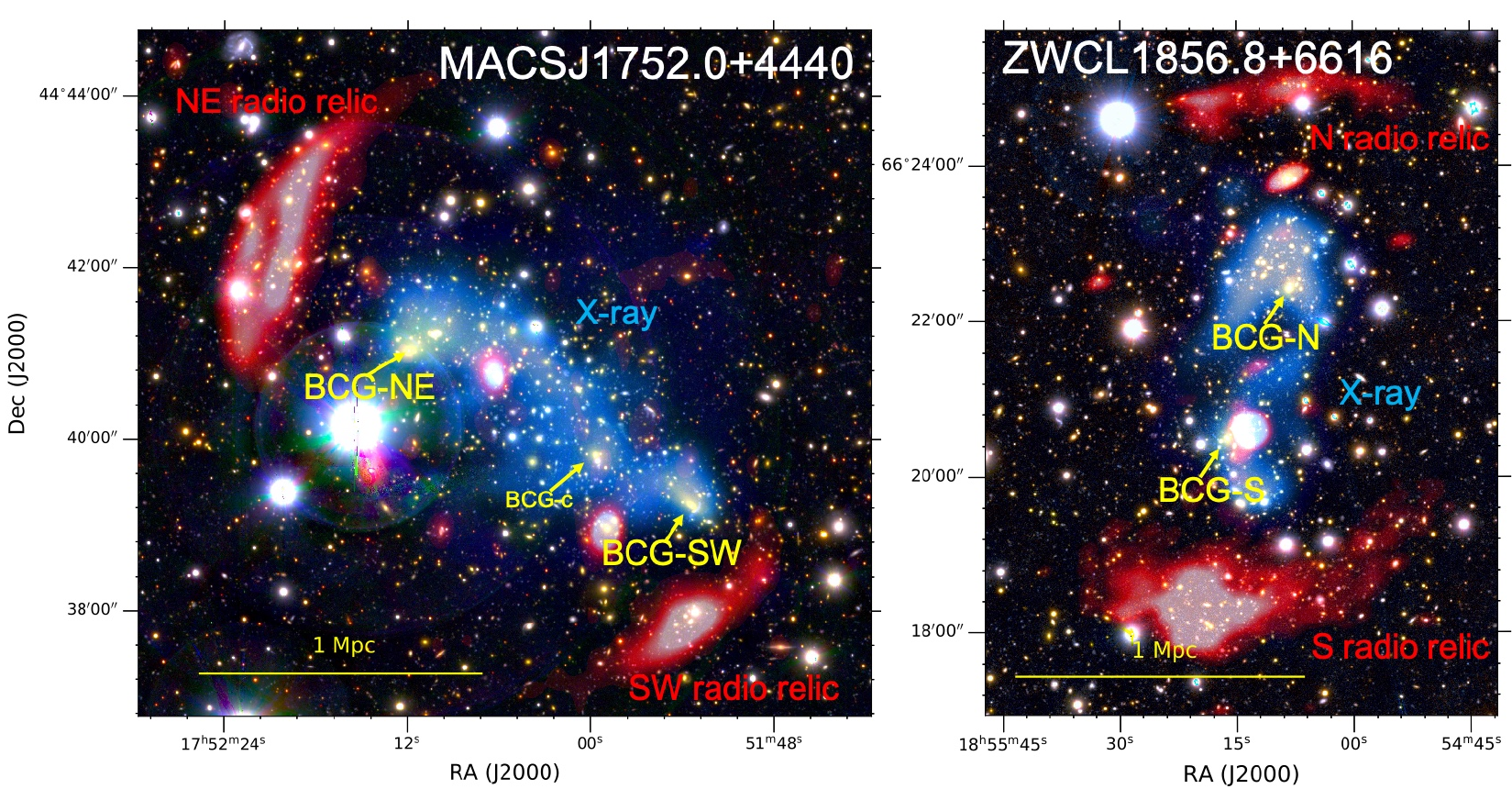}\\
    \caption{Left: Subaru gri color image of MACS1752 with 18cm WSRT radio emission (red) from \cite{2012vanweeren} and 0.5-7~keV XMM-\textit{Newton} X-ray emission (blue). Right: Subaru color image of ZWCL1856 with 144~MHz LOFAR radio emission (red) from Jones et al. in prep. and 0.5-7~keV XMM-\textit{Newton} X-ray emission (blue). The Subaru color image was created by combining g, weighted g+r, and r. Arrows mark the locations of the two brightest cluster galaxies (BCGs) for each cluster. The bright X-ray morphology shows an inverted S shape in both clusters. The two BCGs of each cluster are marked with a yellow arrow. The position and alignment of the radio relics in ZWCL1856 are almost mirrored and are located \edit2{at opposite ends of the elongated X-ray distribution}. In contrast, the relics appear to be misaligned with the ICM distribution in MACS1752.}
    \label{fig:pretty_image}
\end{figure*}

\section{Observations and Data Reduction} \label{sec:data_reduction}
\subsection{Subaru Observations}
\label{sec:subaru_observations}
We observed MACS1752 and ZWCL1856 with the Subaru Telescope Suprime-Cam on the nights of 2013 July 13, 2014 February 26, and 2015 September 12 (PI: D. Wittman). The multi-band imaging is summarized in Table \ref{table:subaru_data}. During observations, a dither and rotation of the camera between exposures was applied to enable minimization of stellar diffraction spikes and removal of saturation trails through our data reduction process. 

Basic data reduction steps (bias and overscan subtraction, flat fielding, geometric-distortion correction, etc.) were performed using the Subaru Suprime-Cam software SDFRED2 \citep{2002yagi, 2004ouchi}. We used SCAMP \citep{bertinscamp} to determine an astrometric solution with Pan-STARRS1 \citep{2016chambers} as the alignment catalog. The astrometric solution was handed to SWARP \citep{bertinswarp} where a median, co-added mosaic image was created. Utilizing the median mosaic image, artifacts such as saturation trails, cosmic rays, bad pixels, etc. were masked and the images were re-SWARPed with the mask weighting to produce a mean, co-added mosaic image. The mean, co-added mosaic images for each galaxy cluster are approximately $40\arcmin$ in diameter and have the standard Suprime-Cam 0\farcs2 pixel scale. Figure \ref{fig:pretty_image} shows our Subaru imaging for MACS1752 (left) and ZWCL1856 (right) with X-ray (blue) and radio observations (red) of the ICM features enhanced in the RGB image.

\begin{table}[]
\caption {Subaru Observations} \label{table:subaru_data}
\def\arraystretch{1.}
\begin{tabular}{ c  c  c }
\hline
\hline
Filter & EXPTIME (s) & DATE-OBS           \\
\hline
\textbf{MACS1752}    &         &                       \\
$g$ & 750     & 2013-07-13             \\
$r$ & 1470    & 2013-07-13            \\
$i$ & 3630    & 2013-07-13/2014-02-26 \\
\hline
\textbf{ZWCL1856} &         &                       \\
$g$  & 750     & 2015-09-12            \\
$r$  & 2190    & 2015-09-12            \\
\hline

\end{tabular}

\end{table}

\subsection{\textit{Hubble} Space Telescope Observations}
\label{sec:hst_obs}
MACS1752 was observed by the \textit{HST} in three programs 12166 (PI: T. Ebeling), 12844 (PI: T. Ebeling), and 13343 (PI: D. Wittman). Since the angular size of the cluster is much larger than the \textit{HST} field-of-view, two pointings, centered on each brightest cluster galaxy (BCG), were obtained. \textit{HST} observations are summarized in Table \ref{table:hst_data}.

We retrieved FLC images from the Mikulski Archive for Space Telescopes (MAST) \footnote{https://mast.stsci.edu/portal/Mashup/Clients/Mast/Portal.html}. These images are delivered pre-calibrated by the STScI OPUS pipeline. Sky subtraction, geometric-distortion correction, and cosmic-ray rejection were achieved with \texttt{Multidrizzle} \citep{2003koekemoer}. To accurately align the \textit{HST} frames, common objects were matched between each frame and the offset was iteratively minimized. The alignment shifts were applied to the images as they were co-added into a mosaic \citep[see][for details]{2014jee}.

\begin{table}[]
\caption {HST/ACS Observations} \label{table:hst_data}
\def\arraystretch{1}
\begin{tabular}{c c c}
\hline
\hline
Filter & EXPTIME (s) & DATE-OBS              \\
\hline
\textbf{MACS1752 NE}    & &  \\
F435W  & 2526     & 2013-10-26            \\
F606W  & 2534    & 2013-10-26            \\
F814W  & 2394    & 2013-10-26            \\
\hline
\textbf{MACS1752 SW}    & & \\
F606W  & 1200    & 2011-08-13            \\
F814W  & 3879    & 2012-12-07/2014-10-26    \\
\hline
\end{tabular}

\end{table}

\subsection{XMM-\textit{Newton} Data Reduction}
Both clusters have XMM-\textit{Newton} X-ray observations. 
MACS1752 was observed on 2010 May 16 (PI: S. Allen; ObsID: 0650383401) for 13 ks and ZWCL1856 was observed on 2013 May 31 (PI: C. Jones; ObsID: 0723160401) for 12 ks.
We retrieved the data from the XMM-{\it Newton} science archive\footnote{http://nxsa.esac.esa.int/nxsa-web/\#home} and reduced them using the XMM-{\it Newton} Extended Source Analysis Software (XMM-ESAS; SAS version 18.0.0; \citep{2004gabriel}).
We followed the procedures described in the XMM-ESAS Cookbook\footnote{http://heasarc.gsfc.nasa.gov/docs/xmm/esas/cookbook/} and conducted standard calibration, filtering, and clearing of flare events.
The final MOS1, MOS2, and PN clean exposure times are 12.2, 12.1, and 6.3 ks for MACS1752 and 11.4, 11.1, and 8.9 ks for ZWCL1856.
For both clusters, we extracted broad band (0.5-7 keV) images for spectral analysis and soft band (0.5-2 keV) images for imaging analysis.
Point sources in the images were detected and removed with the $\tt {edetect\char`_chain}$ task and the resulting holes were filled with the CIAO $\tt {dmfilth}$ task.

\subsection{Point Spread Function Model}
Correction of the image distortion caused by the telescope PSF is a critical aspect of a WL analysis. As the waves of light from the distant galaxies pass through the atmosphere and telescope, the wavefronts are deformed causing a broadening of the observed image. This effect must be modeled and removed in order to access the gravitational-lensing distortions.

The PSF modeling for \textit{HST} and Subaru follow the same basic principles \citep[see][]{2007jee,2017finner}. The objective is to model the PSF of galaxies that will be used to quantify the WL effect. For Subaru obeservations, our pipeline designs PSF models for the galaxies in the mosaic images by modeling the PSF from stars in each of the component frames of the co-added mosaic and then stacking the component PSFs into a co-added PSF. This technique is required because the PSF across the full mosaic image is too complex to be interpolated and the PSF is discontinuous across CCD gaps. For each of the component frames, unsaturated stars were selected by their size and brightness. A principal component analysis (PCA) of the selected stars was performed and the principle components were used to reconstruct position dependent PSF models for each galaxy on a frame-by-frame basis. Finally, the PSF models from the component frames were stacked into a co-added PSF model.

For \textit{HST} data, the PSF modeling has one major difference in that the images of MACS1752 do not contain enough stars to robustly sample the PSF. Instead, following \cite{2007jee}, we relied on archival \textit{HST} images of globular clusters to model the PSF. This technique works because the PSF of the space-based \textit{HST} repeats on its 1.5 hour orbital period. The PSF models\footnote{http://narnia.yonsei.ac.kr/$\sim$mkjee/acs\_psf/} built from the globular cluster images were matched to the MACS1752 images by minimizing the difference between the modeled PSF and the measured shapes of the tens of stars that were found in a single \textit{HST} MACS1752 pointing. Once the best-fit model for each frame was found, the PSF for the galaxies was modeled and co-added, as with Subaru.

\begin{figure}
    \centering
    \includegraphics[width=0.48\textwidth]{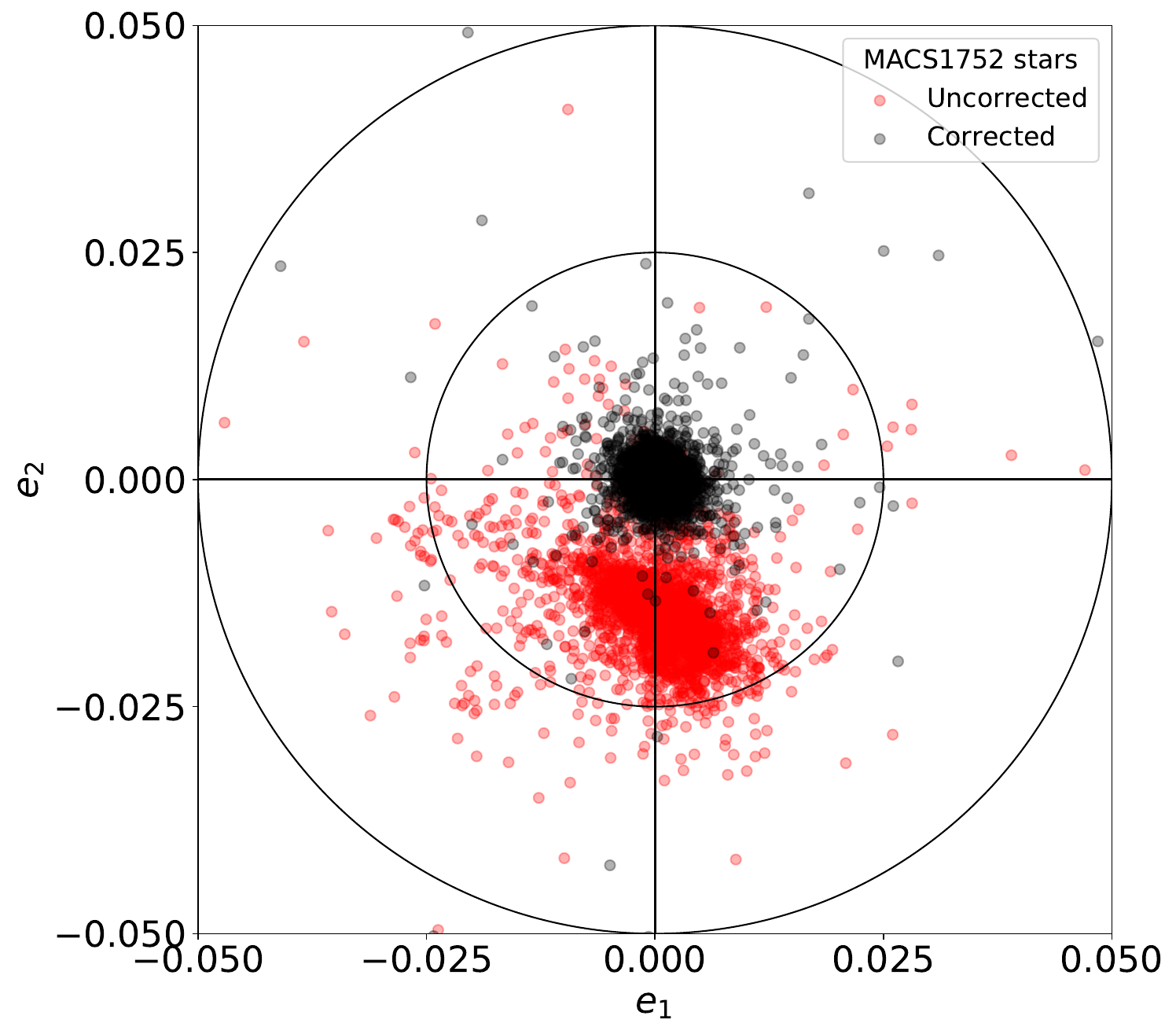}
    \includegraphics[width=0.48\textwidth]{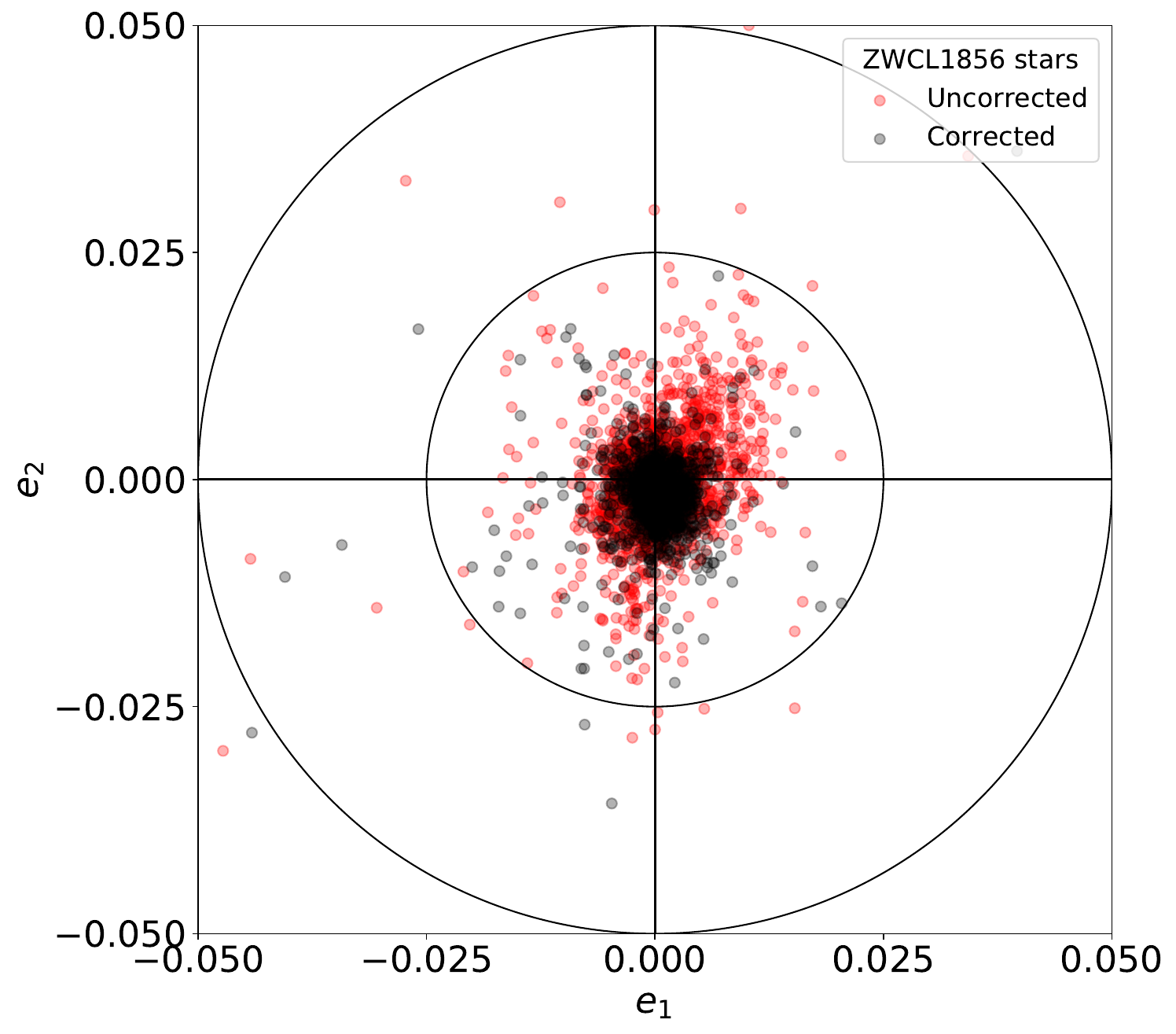}
    \caption{Subaru ellipticity corrections for MACS1752 (top) and ZWCL1856 (bottom). Red circles denote the uncorrected ellipticities of stars from the mosaic image. Black circles are the corrected ellipticities after subtracting the shape of the PSF model for each star. The distributions both tighten and shift towards (0,0).}
    \label{fig:psf_correction}
\end{figure}

In Figure \ref{fig:psf_correction}, we show the ellipticity of stars from the Subaru mosaic images. Red circles mark the star ellipticity components before PSF correction and black circles mark them post correction. As desired, the corrected ellipticities are more circular and centered close to (0,0). The mean values of the corrected ellipticities of stars are $\left<e_1\right>=5\times10^{-4}$ and $\left<e_2\right>=1\times10{-4}$ for MACS1752 and $\left<e_1\right>=5\times10^{-4}$ and $\left<e_2\right>=-1\times10^{-3}$ for ZWCL1856.

\section{Weak-lensing method} \label{sec:wl_method}
\subsection{Theory} \label{sec:theory}
The gravitational potential of a galaxy cluster causes a deflection of the light coming from more distant galaxies. This is observed as a distortion of the shapes of galaxies as described by the following linear transformation matrix:
\begin{equation}
    A = \left( \begin{array}{cc}
 1- \kappa - \gamma_1 & -\gamma_2 \\
 -\gamma_2 & 1 - \kappa + \gamma_1 \\
 \end{array}  \right) \label{eqn_A},
 \end{equation}
which is dependent on $\kappa$, the convergence, and $\gamma$, the shear. The convergence is an isotropic distortion that is a function of the projected mass density $\Sigma$ and the critical surface density $\Sigma_c$:
\begin{equation} \label{eq:kappa}
    \kappa = \frac{\Sigma}{\Sigma_c}. 
\end{equation}
\noindent
The critical surface density 
\begin{equation}\label{eq:sigma_c}
    \Sigma_c = \frac{c^2 D_{s}}{4\pi G D_l D_{ls}},
\end{equation}
is a function of the speed of light $c$, the gravitational constant $G$, and varies with angular diameter distance to the cluster (lens) $D_l$ and with the ratio of angular diameter distances from lens to source and from observer to source $D_{ls}/D_s$. The ratio $\beta = D_{ls} / D_{s}$ is often referred to as the lensing efficiency.
The shear $\gamma$ encodes the anisotropic distortion and can be conveniently described as a complex shear $\gamma = \gamma_1 + i\gamma_2$. In reality, we observe the reduced shear $g_i = \gamma_i / (1 - \kappa)$. Positive and negative values of $g_1$ describe distortions along the x and y axes, respectively. While positive and negative values of $g_2$ describe distortions along the $y=x$ and $y=-x$ directions, respectively. 

\subsection{Shape Measurement}
Detection of the WL effect requires a statistical analysis of the shapes of distorted galaxy images. We employed a model-fitting technique to measure the shapes of galaxies. The shape measurement was performed on the \textit{HST} F814W and Subaru $i$-band imaging for MACS1752 and Subaru $r$-band for ZWCL1856. The procedure for each filter was identical.  

To measure the shape of a single galaxy, we cut out a postage-stamp image of the galaxy from the mosaic image. It is important to consider the size of the postage-stamp image. A large postage stamp will contain the light from nearby objects, which may significantly alter the shape measurement. However, a small postage stamp may prematurely truncate the galaxy light profile and lead to truncation bias (see \cite{2015mandelbaum} for more on truncation bias). We chose to cut large postage-stamp images that are eight times the half-light radius, as measured by \texttt{SExtractor} \citep{1996bertin}, with a 10 pixel floor for very small objects. A corresponding rms-noise postage stamp, $\sigma_{\mathrm{rms}}$, was cut out from the rms mosaic image. Making use of the \texttt{SExtractor} segmentation map, we masked nearby bright objects that could influence the shape measurement by setting the relevant pixels in the rms-noise postage stamp to $10^6$. 

We fit an elliptical Gaussian function, $G$, to the postage-stamp image, $I$, while forward-modeling the corresponding PSF model, $P$, as follows:

\begin{equation}
    \chi^2 = \sum \left(\frac{I - G\circledast P}{\sigma_{\mathrm{rms}}}\right)^2,
    \label{eq:chi2}
\end{equation}
where the summation is over the pixels of the postage stamp. The elliptical Gaussian function has seven free parameters: background, amplitude, position (x and y), semi-major axis ($a$), semi-minor axis ($b$), and orientation angle ($\phi$). We fixed the background, $x$, and $y$ while fitting the remaining four parameters. Complex ellipticities were determined from $a$, $b$, and $\phi$ as

\begin{align}
e_1 &= \frac{a-b}{a+b}\cos 2\phi, \\
e_2 &= \frac{a-b}{a+b}\sin 2\phi.
\end{align}
This procedure was repeated for each galaxy in each filter and filter-dependent shape catalogs were compiled. It is known that galaxy shape measurement techniques suffer from bias \citep[for more on WL biases see][]{2018mandelbaum}. We derived a calibration factor from simulations of our pipeline to correct for multiplicative bias. The multiplicative calibration factors are 1.15 (1.25) for Subaru (\textit{HST}) ellipticities. These calibrations were derived through pipeline tests on simulated images \citep{2013jee}. At this point, the catalog contains stars as well as foreground, cluster, and background galaxies. In the following section, the shape catalogs will be refined to a source catalog containing the lensed background galaxies.

\subsection{Source Selection} \label{sec:source_selection}
The ideal catalog for a WL analysis contains only galaxies that are at a greater distance than the lens (cluster) from the observer. Direct distance measurements for each object would make source selection trivial. As one can imagine, gathering spectroscopic redshifts for of order 20,000 galaxies is an immense undertaking. Photometric redshifts rely on multi-band imaging, which is often not available for specific galaxy clusters. With redshifts unavailable, we instead rely on the color and magnitude of galaxies to categorize them. The rest-frame 4000\AA\ break, a prominent feature in the spectral energy distribution (SED) of evolved cluster galaxies, forms a red-sequence relation in color when observed with two filters that bracket the feature. At the redshifts of ZWCL1856 (0.304) and MACS1752 (0.365), the 4000\AA\ breaks are at wavelengths of 5216\AA\ and 5460\AA, respectively. The feature is well-bracketed by the $g$ and $r$ or $g$ and $i$ filters.  

\begin{figure}
    \centering
    \includegraphics[width=0.5\textwidth]{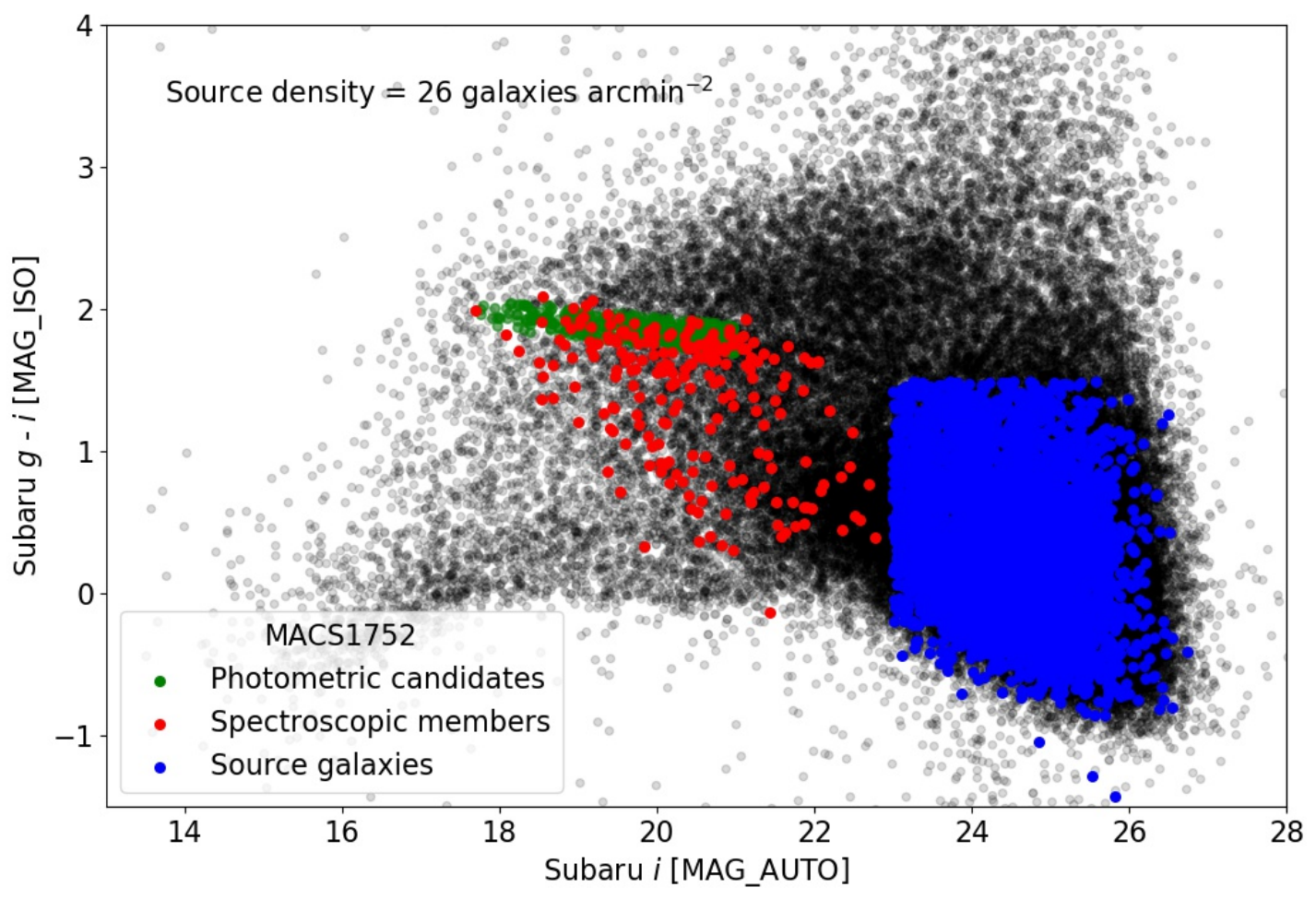}\\
    \includegraphics[width=0.5\textwidth]{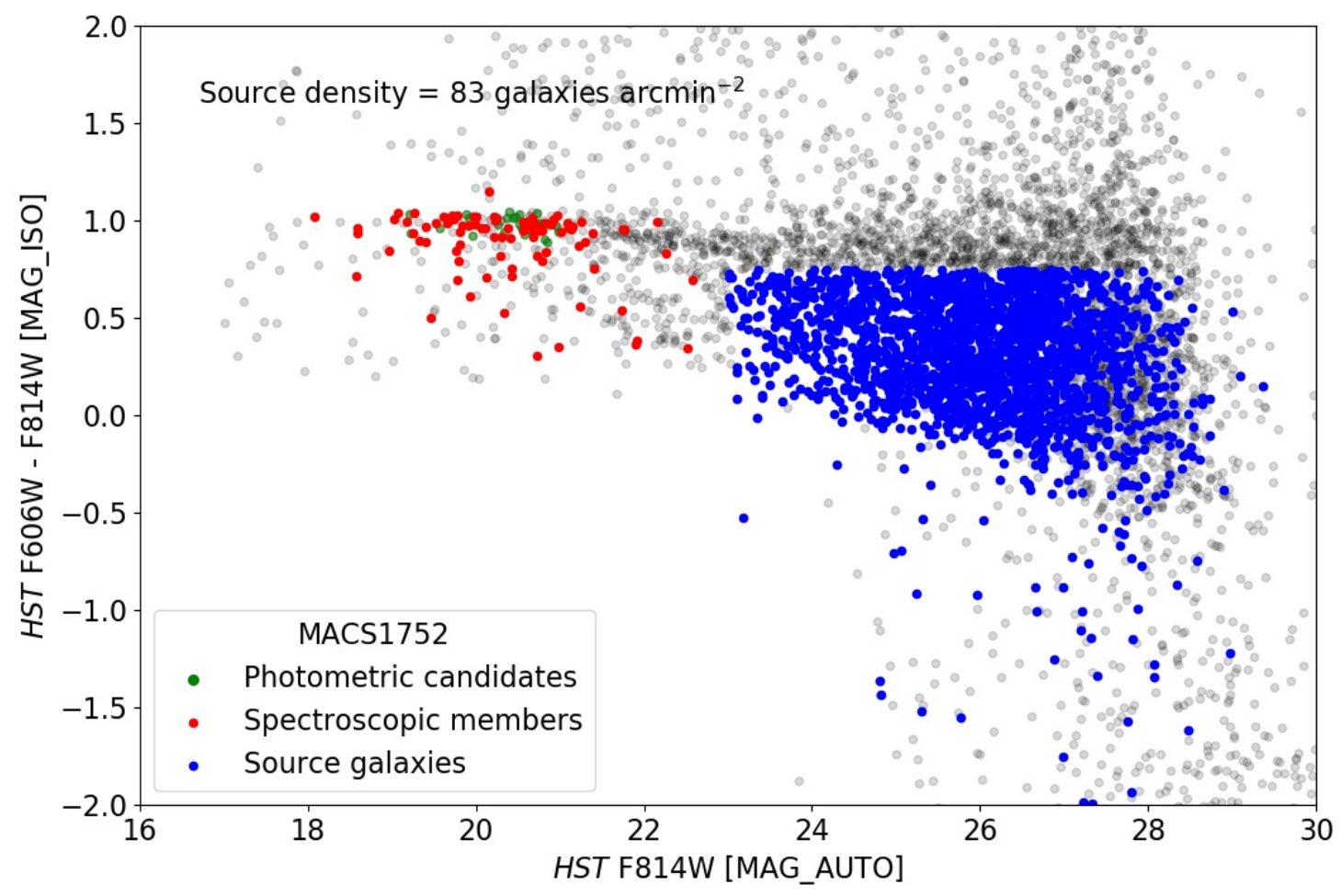}
    \caption{Color-magnitude diagram for all detected sources (black circles) in the MACS1752 Subaru (top) and \textit{HST} (bottom) imaging. Red circles represent spectroscopically confirmed cluster members within redshift of 0.03 of the BCG. A tight red sequence relation is found in the \textit{HST} CMD. Extending the range of spectroscopic cluster members to the Subaru field of view shows that there are many bluer members. By a linear fit to the reddest spectroscopic member galaxies (the red sequence), galaxies within $(g-i)\pm0.1$ in Subaru or $(F606W-F814W)\pm0.1$ in \textit{HST} were selected as photometric red-sequence candidates (green circles). Blue circles are the galaxies selected for the background source catalog based on their relation to the red sequence and the shape criteria mentioned in the Section \ref{sec:source_selection}. The dense feature found underneath the red sequence in the Subaru CMD arises from Galactic stars and shows a trend of brighter and bluer.}
    \label{fig:cmd_macs}
\end{figure}

For MACS1752, we modeled shapes in \textit{HST} and Subaru imaging. Figure \ref{fig:cmd_macs} displays the color-magnitude diagrams (CMDs) for all detected sources (black circles) in the MACS1752 Subaru imaging (top) and \textit{HST} imaging (bottom). The red circles represent galaxies with spectroscopic redshifts from the catalog of \cite{2019agolovich} that are within $\Delta z=0.03$ of the BCG redshift. We set the criteria for background galaxies to be those that are bluer than the red sequence and fainter than 23rd magnitude. Furthermore, we required the pre-seeing semi-minor axis of objects to be $b > 0.3$ pixels to remove objects that are too small to be lensed galaxies, the magnitude uncertainty to be $dm < 0.3$, and the measured ellipticity $e < 0.9$ and its uncertainty $de < 0.3$ to prevent poorly fit objects from entering the source catalog. The $b$ and $de$ criteria are effective for star-galaxy separation because the pre-seeing light profile of stars is close to a delta function, which has no size and returns a poor shape fit. These selection criteria prevent stars and other spurious objects from entering the source catalog. This source selection provides \mytilde 26 galaxies arcmin$^{-2}$ (\mytilde 83 galaxies arcmin$^{-2}$) for MACS1752 Subaru (\textit{HST}) imaging.

\begin{figure}
    \centering
    \includegraphics[width=0.5\textwidth]{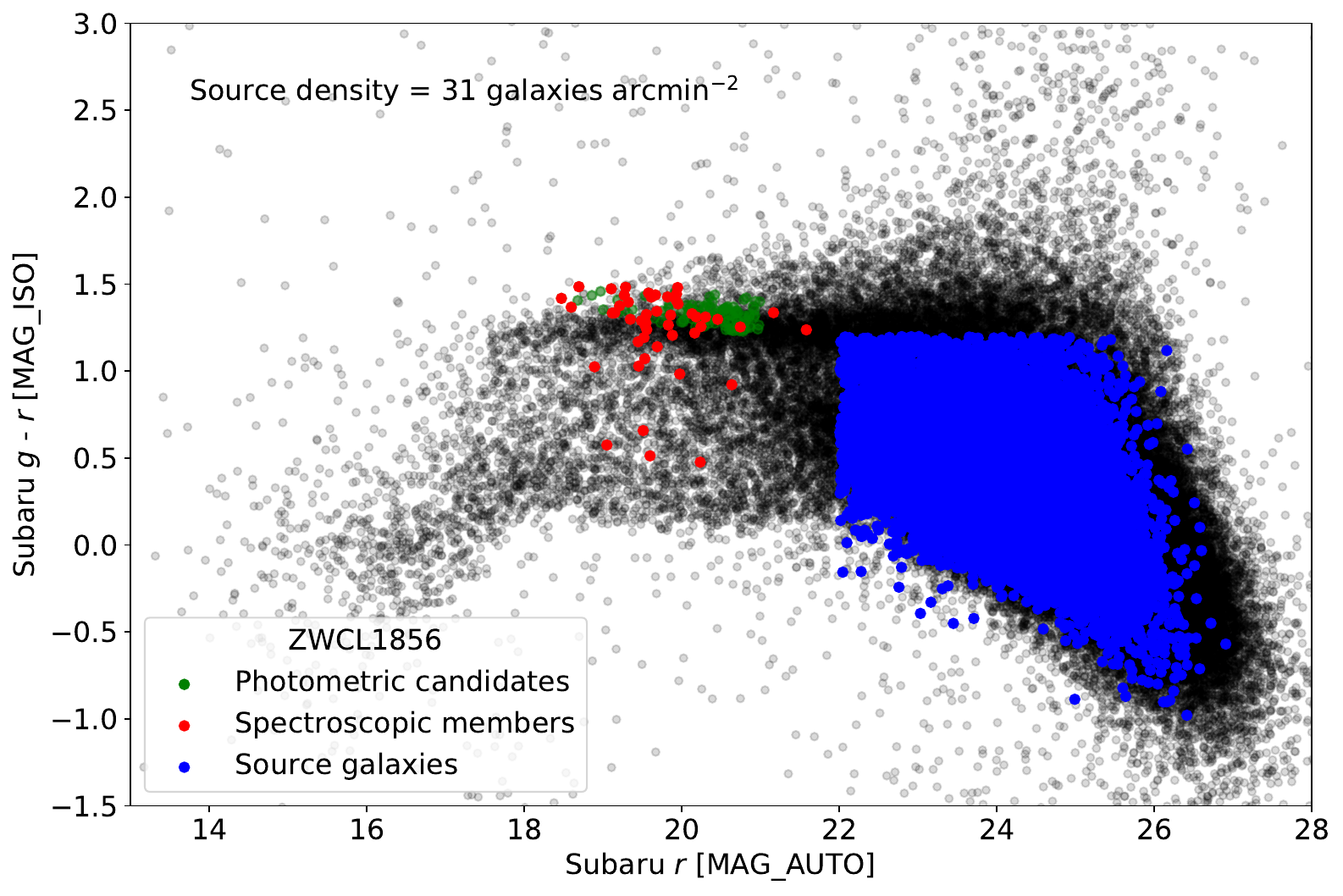}
    \caption{Color-magnitude diagram for all detected sources (black circles) in the ZWCL1856 Subaru field of view. Red circles are spectroscopically confirmed cluster members. Green circles are the photometric red-sequence candidate selection. Blue circles represent galaxies selected for the background source catalog. The dense feature found underneath the red sequence arises from Galactic stars and shows a trend of brighter and bluer.}
    \label{fig:cmd_ZWCL_sub}
\end{figure}

Figure \ref{fig:cmd_ZWCL_sub} shows the CMD for all detected sources (black circles) in the ZWCL1856 imaging. Source galaxies (blue circles) in ZWCL1856 were selected using the same constraints as for the Subaru imaging of MACS1752, except that the magnitude cut was set to the 22nd magnitude to take advantage of the lower redshift of the cluster. The source density is \mytilde 31 galaxies arcmin$^{-2}$ for ZWCL1856 Subaru imaging. 

\subsection{Source Redshift Estimation}
In a WL analysis, each source galaxy provides a probe of the projected galaxy cluster potential. As is apparent in Equation \ref{eq:sigma_c}, the effectiveness of the gravitational lens varies on a source-by-source basis with the lensing efficiency, $\beta$. To remedy the absence of distance measurements for background galaxies, we utilize the photometric redshift catalog \citep{2010dahlen} of the GOODS-S field as a reference for our source catalog. This technique is common in WL studies that do not have the luxury of redshifts for each source galaxy \citep[to name a few, ][]{2011jee, 2016okabe, 2018schrabback}.

For each of the three source catalogs, a GOODS-S reference catalog was derived by applying the same magnitude and color constraints as was applied during source selection (Section \ref{sec:source_selection}). Photometric transformations were applied when necessary to convert the \textit{HST} filters to match the Subaru filters. Then, the galaxy number density as a function of magnitude for each source catalog was compared to the corresponding constrained GOODS-S reference catalog. The difference in depth was corrected by weighting the constrained GOODS-S reference catalog to match the number density of each source catalog. Finally, the three constrained and weighted GOODS-S photometric redshift distributions were used to estimate the lensing efficiency and corresponding effective redshift of the galaxies in our source catalogs. The lensing efficiency was restricted to
\begin{equation}
    \beta = \left<max\left(0,\frac{D_{ls}}{D_s}\right)\right>,
\end{equation}
which forces foreground galaxies to have $\beta=0$. The lensing efficiency of Subaru source galaxies for MACS1752 is $\left<\beta\right>=0.53$ (effective redshift $z_{\mathrm{eff}}=0.89$) with a width of the distribution of $\left<\beta^2\right>=0.33$ and for \textit{HST} $\left<\beta\right>=0.62$, $z_{\mathrm{eff}}=1.18$, and $\left<\beta^2\right>=0.41$. The lensing efficiency for ZWCL1856 is $\left<\beta\right>=0.54$ with $z_{\mathrm{eff}}=0.76$ and width of the distribution of $\left<\beta^2\right>=0.36$. 

The bias caused by approximating the source distance and distribution of the source population by a single characteristic value has been discussed in detail in \cite{1997seitz} and \cite{2000hoekstra}. In our analysis, the width of the distribution is accounted for by modifying the reduced shear $g$ to the corrected reduced shear:

\begin{equation}
g' = \left[1 + \left(\frac{\left<\beta^2\right>}{\left<\beta\right>^2}-1\right) \kappa \right] g,
\end{equation}
where $\kappa$ is the convergence as defined in Section \ref{sec:theory}.
\section{Results} \label{sec:results}


\subsection{Mass Reconstruction} \label{sec:wl_analysis}

The Subaru observations have a footprint that extends approximately 40 arcmins in diameter. This footprint covers $3\sim4$ virial radii at the redshifts of ZWCL1856 ($z$=0.304) and MACS1752 ($z$=0.365). 
We used the FIATMAP code \citep{1997fischer}, which performs a real-space convolution, to convert galaxy ellipticities to a convergence map 

\begin{figure*}
    \includegraphics[width=1.\textwidth]{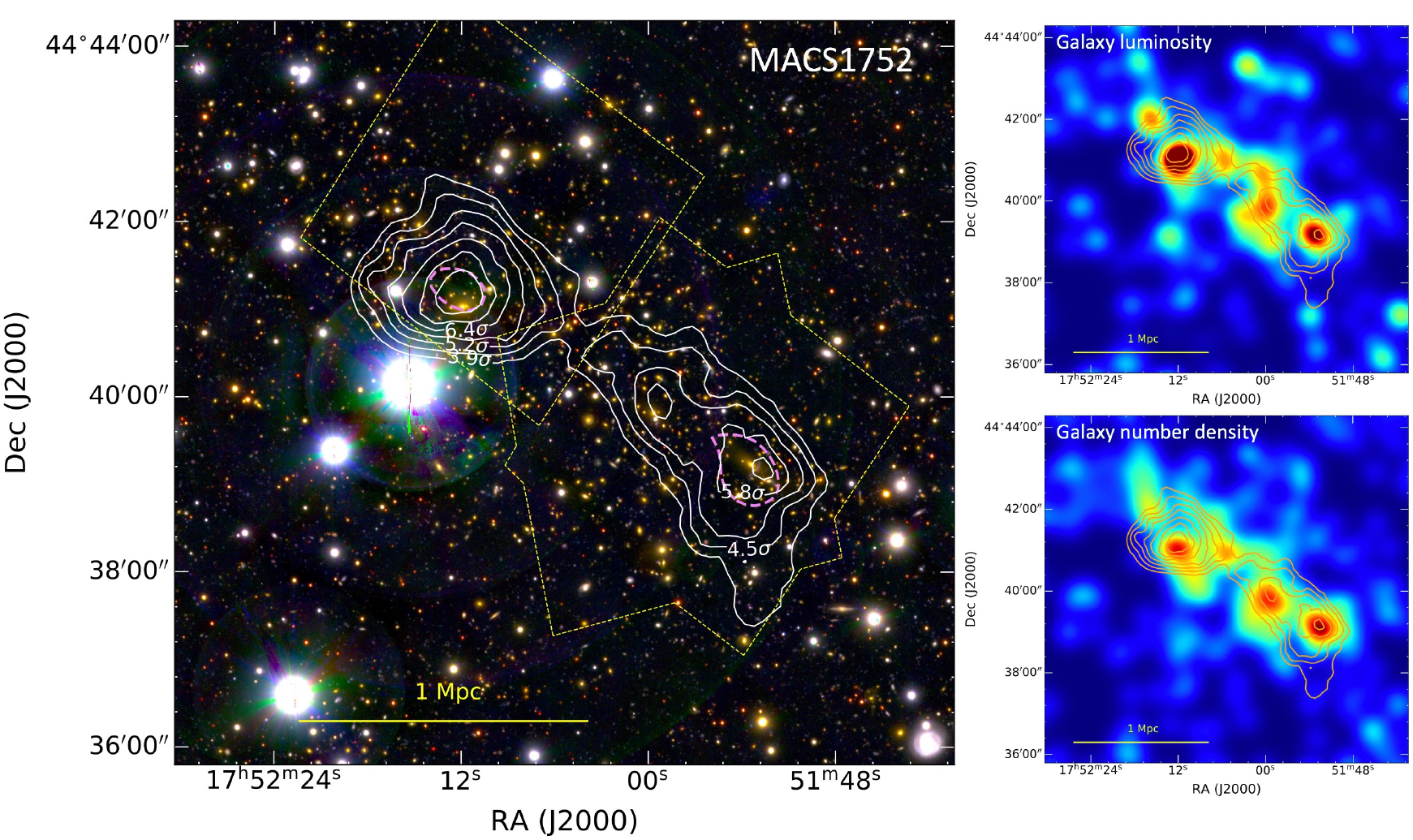}
    \caption{Left: Fiatmap reconstruction of the mass distribution of MACS1752 derived from \textit{HST} and Subaru shapes plotted over Subaru color image. Contours are labeled with S/N values derived from bootstrapping. The dashed violet contours show the 1$\sigma$ peak uncertainty from bootstrapping. The mass peaks are in statistical agreement with the respective BCG. The dashed yellow line outlines the \textit{HST} footprint. Top-right: Convergence contours (orange) over the colormap of $i$-band luminosity of cluster galaxies. Bottom-right: Convergence contours (orange) over the colormap of number density of cluster galaxies. Galaxy distributions are smoothed with a $\sigma=18''$ Gaussian kernel. }
    \label{fig:macsj1752_fiatmap}
\end{figure*}

The mass distribution of MACS1752 is shown in the left panel of Figure \ref{fig:macsj1752_fiatmap}. Two dominant mass peaks are detected at opposite ends of the mass distribution that have excellent agreement with the NE and SW BCGs. In addition, a substructure is detected between the primary clusters, \mytilde 443 kpc northeast of the SW BCG. This mass substructure is co-spatial with the third BCG. To test the significance of the peaks, we performed bootstrap resampling to create 1000 realizations of the source catalog and mass map. The resulting $1\sigma$ distributions of the mass peaks are marked as the violet dashed contours in the left panel of Figure \ref{fig:macsj1752_fiatmap}. The signal-to-noise (S/N) contours derived from bootstrapping are labeled in Figure \ref{fig:macsj1752_fiatmap} and the peak S/N values are summarized in Table \ref{table:mass_estimations}. The NE mass peak has the strongest detection at the 7.9$\sigma$ level, followed by the SW mass peak at 6.5$\sigma$ and finally the central peak at $5.5\sigma$. 

The \cite{2019bgolovich} spectroscopic analysis of MACS1752 identified 176 cluster galaxies (red circles in Figure \ref{fig:cmd_macs}). We supplement the cluster galaxies by selecting photometric candidates from the CMD within Subaru $(g-i)\pm0.1$ or \textit{HST} $(F606W-F814W)\pm0.1$ along a linear fit to the red sequence. This selection is highlighted by the green circles in Figure \ref{fig:cmd_macs}. Combining the two catalogs of cluster galaxies, we plot the luminosity and number density distributions in Figure \ref{fig:macsj1752_fiatmap}. The S/N contours trace the luminosity (top-right panel) and number density (bottom-right panel) distributions of the cluster galaxies. As expected, the brighter peaks and higher number density coincide with the S/N peaks.

\begin{figure*}
    \includegraphics[width=0.8\textwidth]{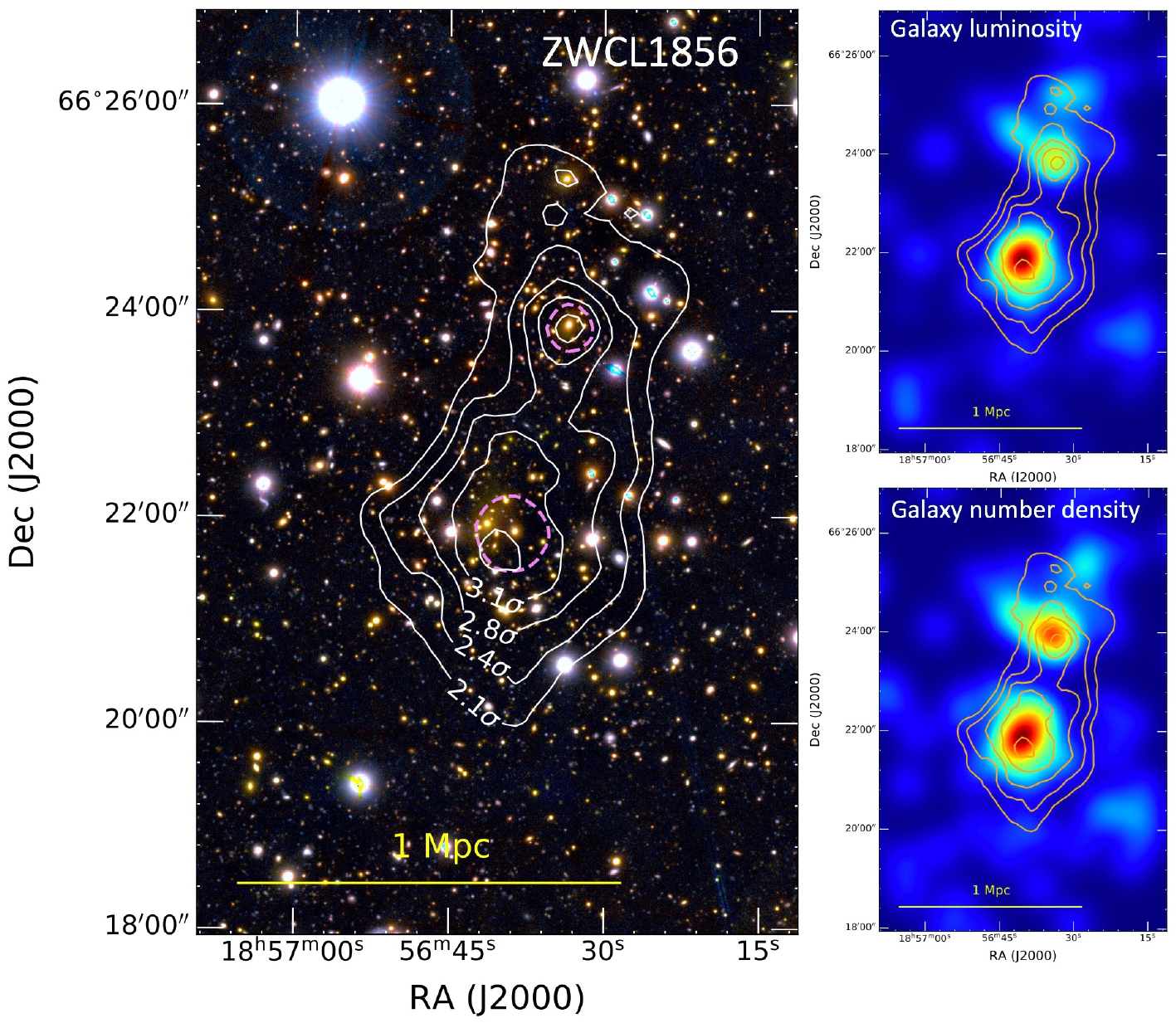}
    \caption{Fiatmap reconstruction of the mass distribution of ZWCL1856 with Subaru color image in background. The dashed violet contours show the 1$\sigma$ peak uncertainty from bootstrapping. Top-right: Convergence contours (orange) plotted over the $r$-band luminosity of cluster galaxies. Bottom-right: Convergence contours (orange) over the number density of cluster galaxies. Galaxy distributions are smoothed with a $\sigma=18''$ Gaussian kernel.}
    \label{fig:ZWCL_massmap}
\end{figure*}

The mass map of ZWCL1856 (Figure \ref{fig:ZWCL_massmap}) shows two dominant peaks. From bootstrapping, the peak significances are S/N=$4\sigma$ in the north and S/N=$3.7\sigma$ in the south. Both peaks show excellent agreement with the corresponding BCG locations. An elongation from the southern mass peak to the center is detected that is similar to that in MACS1752 and, although there are some cluster galaxies in the region, none are comparable in brightness to the north and south BCGs. The spectroscopic coverage of ZWCL1856 is more sparse (47 cluster member galaxies) than MACS1752. Photometric candidates (green circles in Figure \ref{fig:cmd_ZWCL_sub}) were selected within ($g-r)\pm0.1$ of a linear fit to the spectroscopically-defined red sequence.  The luminosity distribution (top-right panel) reveals that the southern peak is 1.5 times brighter than the northern, which is in contrast to the lensing S/N that is 1.1 times higher for the northern peak. However, the number density distribution (bottom-right panel) shows a more equal weight to each subcluster.

\subsection{Mass Estimation}
The lensing signal that is detected from merging galaxy clusters arises from the entire mass distribution. Therefore, to properly determine the mass from the lensing signal, the contribution of each structure to the observed shear should be carefully modeled. Our convergence maps reveal that MACS1752 is composed of three significant subclusters and ZWCL1856 is composed of two. In this section, we derive the masses of these subclusters. 

We estimated the mass of each subcluster by superposing NFW  \citep{1997navarro} halos and modeling the expected tangential shear at each source-galaxy position. The radial dependence of the shear was calculated based on the equations from \cite{2000wright}. We fixed the centroid of each subcluster halo to its respective BCG. Due to recent merging activity, the BCG may not necessarily coincide with the mass peak \citep[e.g.][]{2014martel}. However, for the two clusters that are presented in this work, the mass peaks and their corresponding BCGs are in good agreement. Nevertheless, the core regions (\mytilde 200 kpc) around each subcluster were removed to bypass the effects of the strong-lensing regime and the region where additional systematic effects can lead to a biased mass estimate. We elected to not use a concentration-mass ($c$-$M$) relation to estimate the mass of each subcluster because most $c$-$M$ models are valid only within the cosmology and mass range of the simulations from which they are derived. Instead, we sampled the posterior distribution of the mass and concentration with Markov Chain Monte Carlo (MCMC). Priors were set on the concentration and mass of $1 < c < 6$ and $10^{13}\ $M$_\odot < M_{200} < 10^{16}\ $M$_\odot$. These priors amply cover the typical range of galaxy cluster masses and concentrations that arise in numerical simulations \citep[e.g.][]{2008duffy, 2019diemer}. The mass estimates from MCMC are summarized in Table \ref{table:mass_estimations} and reveal that both MACS1752 and ZWCL1856 are major mergers with approximately 1:1 mass ratio. Concentrations are unconstrained by this method. Since the central subcluster in MACS1752 is low mass ($0.3^{+0.4}_{-0.1}\times10^{14}\ $M$_\odot$), we exclude it from our examination of the merger.

The total $M_{200}$ of each system is not simply the addition of the subcluster masses but is the mass within a radius at which the combined average density of the subclusters is 200 times the critical density of the universe at the cluster redshift. To calculate the total mass, two NFW halos were stacked in a 3D grid with locations given by their measured projected separation. Integrating radially from the midpoint, the total mass was calculated. This procedure was repeated for each sample from the MCMC chains to sample the total mass distribution without using a $c$-$M$ relation. The total mass of MACS1752 is $M_{200}=14.7^{+3.8}_{-3.3}\times10^{14}\ $M$_\odot$ and the total mass of ZWCL1856 is $M_{200}=2.4^{+0.9}_{-0.7}\times10^{14}\ $M$_\odot$. The total mass of MACS1752 is higher but $1\sigma$ consistent with the mass estimated from the SZE. However, the total mass of ZWCL1856 is lower and inconsistent with its mass estimated from the SZE. This is not surprising because mass estimates from the SZE rely on the state of the ICM, which should not be considered to be in hydrostatic equilibrium for a merging system. Furthermore, treating a system that consists of two subclusters as a single cluster could subject the SZE mass estimate to additional bias.






\begin{table*}[h!t]

\caption {Measured Properties of MACS1752 and ZWCL1856} \label{table:mass_estimations}
\def\arraystretch{1}
\begin{tabular}{c c c c c c c c c}
\hline
\hline
Subcluster & RA & Dec. & BCG redshift & $M_{200}$ & Peak S/N  & $M/L$ & $\sigma_v$ & $T$        \\
& & & & $10^{14}  $M$_\odot$ & & M$_\odot$ L$^{-1}_\odot$ & km s$^{-1}$ & keV \\
\hline
\textbf{MACS1752}    & & &  & $^{a}$ & & $^{b}$\\
 
NE & $17^\mathrm{h}52^\mathrm{m}11^\mathrm{s}.9$ & $44^\circ41\arcmin02\arcsec$ & 0.3648 & $5.6^{+1.8}_{-1.6}$  & $7.9$ & $498^{+102}_{-100}$ & $1006\pm60$ & $7.6^{+2.4}_{-1.5}$             \\
SW & $17^\mathrm{h}51^\mathrm{m}53^\mathrm{s}.4$ & $44^\circ39\arcmin14\arcsec$ & 0.3634 & $5.6^{+1.4}_{-2.1}$  & 6.5 & $412^{+91}_{-93}$ & $1038\pm77$ & $6.8^{+3.4}_{-1.7}$          \\
Center & $17^\mathrm{h}51^\mathrm{m}59^\mathrm{s}.6$ & $44^\circ39\arcmin49\arcsec$ & 0.3617 & $0.3^{+0.4}_{-0.1}$    & 5.5            \\
\hline
\textbf{ZWCL1856}    & &\\
N & $18^\mathrm{h}56^\mathrm{m}33^\mathrm{s}.6$ & $66^\circ23\arcmin57\arcsec$ & 0.3041 & $1.2^{+0.5}_{-0.5}$     & 4.0 & $360^{+102}_{-106}$ & $934\pm117$ & $4.3^{+2.1}_{-1.1}$            \\
S & $18^\mathrm{h}56^\mathrm{m}41^\mathrm{s}.5$ & $66^\circ21\arcmin56\arcsec$ & 0.3033 & $1.0^{+0.4}_{-0.7}$    & 3.7 & $183^{+84}_{-97}$ & $862\pm133$ & Unconstrained     \\
\hline
\end{tabular}
\begin{tablenotes}
    \small
    \item $^{a}$Combined Subaru and \textit{HST} sources boost S/N for MACS1752
    \item $^{b}$Velocity dispersion ($\sigma_v$) measurements from \cite{2019bgolovich}
\end{tablenotes}
\end{table*}

\subsection{$M/L$ Ratio}
The mass-to-light ratio ($M/L$) of galaxy clusters has been used as a probe of cosmology to measure the matter density of the universe. \cite{1996carlberg} examined sixteen X-ray luminous galaxy clusters and determined the $M/L$ ratio of clusters is \mytilde300 M$_\odot$/L$_\odot$ by using the velocity dispersion of cluster galaxies. \cite{2002girardi} showed that the $M/L$ ratio of galaxy clusters is positively correlated with cluster mass and has a large scatter that ranges from $50 \sim 1000$ M$_\odot$/L$_\odot$. 

We measured the $M/L$ ratio for each subcluster of MACS1752 and ZWCL1856. Centered on the respective BCG, equal-sized apertures were placed that maximize the size without overlap (MACS1752 radius=550 kpc, ZWCL1856 radius=300 kpc). These apertures were used to evaluate the projected mass and galaxy luminosity within.  The projected mass was calculated by integrating the projected density of an NFW halo \citep{2000wright} that was modeled with our WL mass and concentration samples from MCMC. To derive the total luminosity within the aperture, we used our cluster galaxy catalogs (compiled in Section \ref{sec:wl_analysis}) that include the spectroscopically confirmed cluster members and the photometrically selected sample. To ease comparison with literature, $i$- (MACS1752) and $r$-band (ZWCL1856) magnitudes were converted to absolute magnitudes and then transformed to Johnson $B$-band absolute magnitudes by synthetic photometry \citep{2005sirianni} with SED templates for elliptical, spiral, and irregular galaxies. Lastly, solar luminosities were calculated from the magnitudes by assuming $B_\odot=5.48$. The $M/L$ ratios are presented in Table \ref{table:mass_estimations} and are within the range of $M/L$ ratios that are predicted from \cite{2002girardi}. Furthermore, the $M/L$ ratios follow the expected trend with more massive clusters having larger ratios.

\subsection{X-ray analysis} \label{sec:xray_analysis}
X-ray analysis plays an important role in interpreting merger scenarios, especially in radio relic clusters. To our knowledge, no in-depth analysis of the archival X-ray data on these specific clusters has been published.

\begin{figure*}
    \includegraphics[width=1\textwidth]{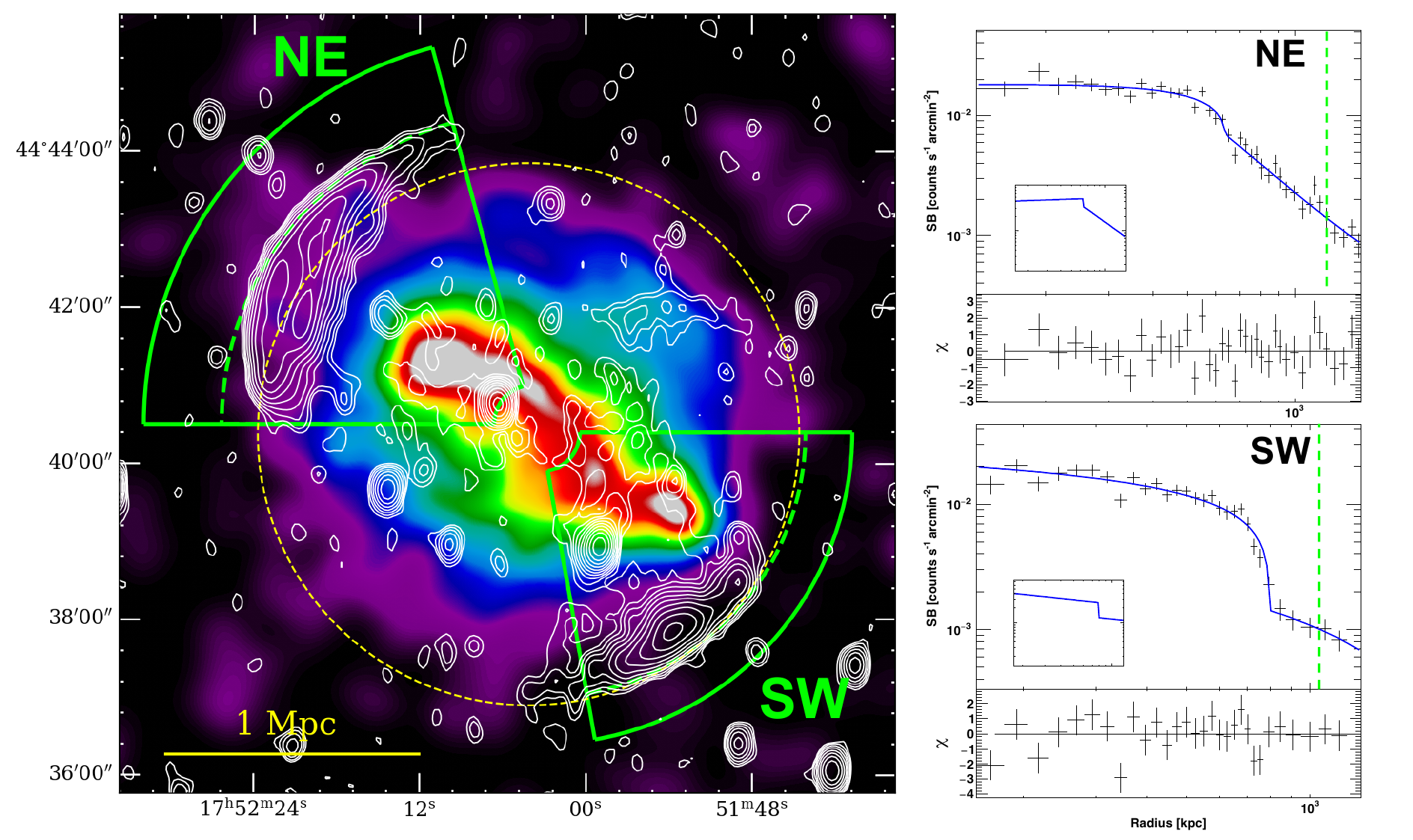}
    \caption{Left: Adaptively smoothed, background-subtracted, exposure-corrected XMM-{\it Newton} image for MACS1752. Green fan-shaped regions correspond to the areas for the X-ray surface brightness analysis shown in right panels. The yellow dashed circle indicates the region used to extract the global X-ray temperature. Top right: X-ray surface brightness profile (black cross) with a line of sight projected and XMM-{\it Newton} PSF convolved broken power-law fit (blue solid line) for the northeast core. The edge of the relic is indicated by the green dashed line. Inset image shows the 3D gas density model. Bottom right: Same as the upper panel, but for the southwest core.}
    \label{fig:xray_macs}
\end{figure*}

\begin{figure*}
    \includegraphics[width=1\textwidth]{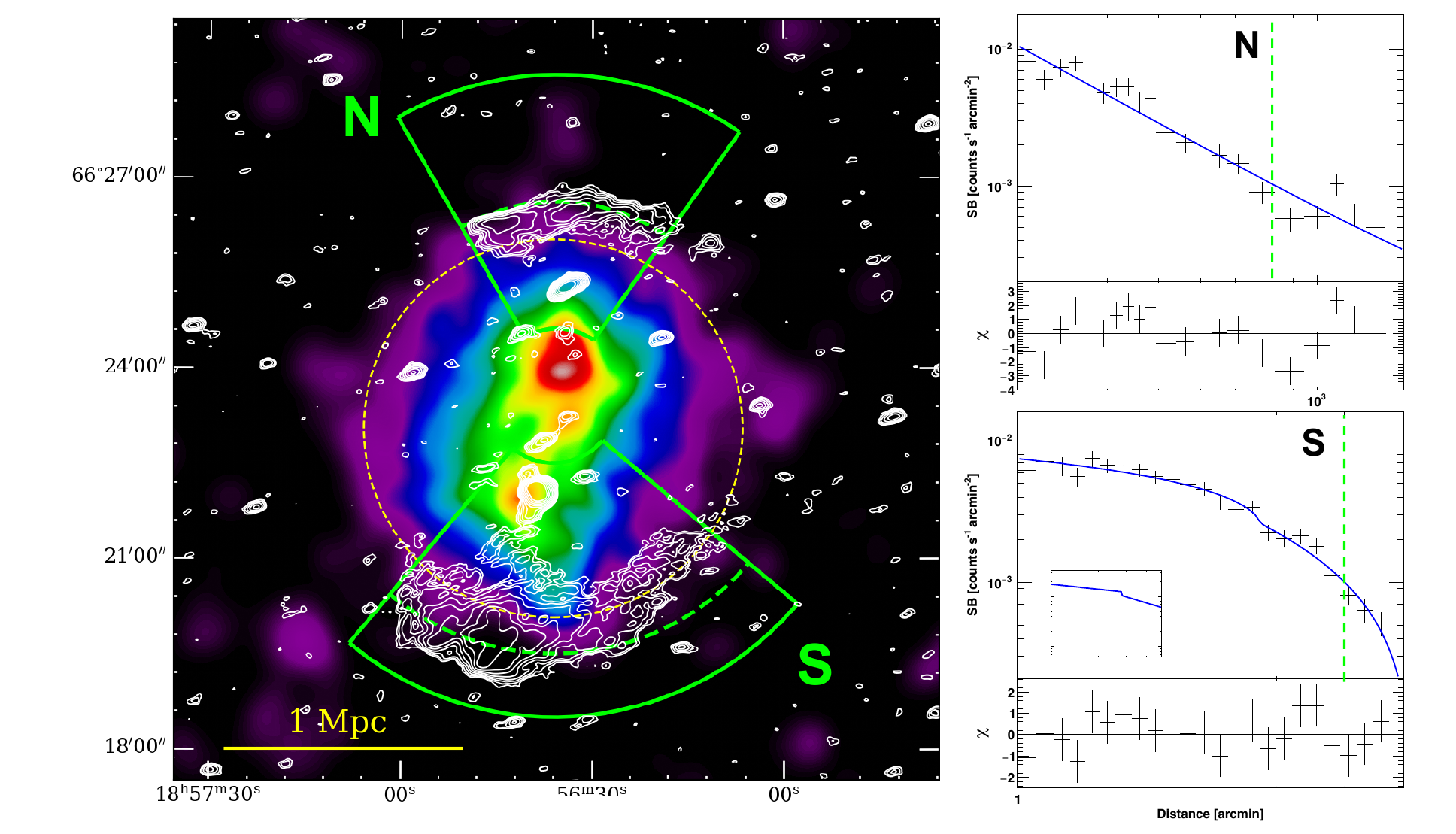}
    \caption{Left: Adaptively smoothed, background-subtracted, exposure-corrected XMM-{\it Newton} image for ZWCL1856. Green fan-shaped regions correspond to the areas for the X-ray surface brightness analysis shown in right panels. The yellow dashed circle indicates the region used to extract the X-ray temperature. Top right: X-ray surface brightness profile (black cross) with a line of sight projected and XMM-{\it Newton} PSF convolved power-law fit (blue solid line) for the northern core. The edge of the relic is indicated by the green dashed line. \edit2{Bottom right: Same as the upper panel but with broken power-law fit for the southern subcluster}. 
    }
    \label{fig:xray_ZWCL}
\end{figure*}

To measure average global temperatures of the systems, we extracted spectra from the point-source-masked MOS1, MOS2, and PN images within a circular region $\mytilde6\arcmin$ in diameter (yellow dashed circles in Figures \ref{fig:xray_macs} and \ref{fig:xray_ZWCL}). The net photon counts (net background) from MOS1, MOS2, and PN for MACS1752 are 4050, 3767, 8082 (85, 84, 136) and for ZWCL1856 are 1957, 1930, 6874 (57,54,122). The combined MOS and PN spectra were analyzed using $\tt{XSPEC}$ ($\tt{v12.10.1f}$) with the $\tt {APEC}$ plasma model and the $\tt {PHABS}$ photoelectric absorption model in the broad energy band (0.5-5~keV).
We assumed a fixed abundance of $0.3~Z_{\odot}$ and fixed neutral hydrogen column densities of $3.2\times10^{20}~\mbox{cm}^{-2}$ and $9.9\times10^{20}~\mbox{cm}^{-2}$ for MACS1752 and ZWCL1856, respectively \citep{2016hipass}.
We find the global temperatures of MACS1752 and ZWCL1856 to be $kT=5.9^{+1.0}_{-0.9}~\mbox{keV}$ and $kT=3.7^{+0.6}_{-0.5}~\mbox{keV}$. The same method was used to determine temperatures for each subcluster within the apertures used for $M$/$L$ determination. These temperatures are presented in Table \ref{table:mass_estimations}. The temperature of ZWCL1856 S was unable to be constrained due to low source counts.



%
%

As radio relics are expected to arise at merger shocks, we attempted to detect the shock in the X-ray emission from the ICM. Using the radio relic locations as a guide, we extracted the X-ray surface brightness profile from the soft-band (0.5-2~keV) background-subtracted, exposure-corrected, point-source-subtracted images in fan-shaped regions marked with green lines in Figures \ref{fig:xray_macs} and \ref{fig:xray_ZWCL}. 
The extracted profiles were rebinned to have a minimum S/N $>$ 5. Using $\tt {PROFFIT~v1.5}$ \citep{2011eckert}, we first attempted to fit a broken power law to the extracted surface brightness profile and if that failed we fit a single power law.

The right panels of Figure \ref{fig:xray_macs} show a surface brightness drop between the leading edge of both MACS1752 subclusters and their respective radio relic with compression factors of $C=1.5\pm0.2$ and $C=2.6\pm0.3$ for the NE and SW subclusters, respectively. We interpret these surface brightness drops as indicators of cold fronts (contact discontinuities) because they do not coincide with the radio relics.

We performed the same analysis for ZWCL1856 and found that a single power-law fit best with no visible drop in the north (Figure \ref{fig:xray_ZWCL}). A double power-law fit in the south finds a drop in surface brightness with a compression factor of $C=1.2\pm0.1$ but not coincident with the radio relic. 

\edit2{We attempted to determine the temperature of the region outside the surface brightness drops for both clusters but a robust detection was not possible because of the short exposure observations. For both clusters, additional X-ray observations will be required to explore the vicinity of the radio relics.}


\section{Discussion} \label{sec:discussion}
\subsection{Identification of the Substructures}
The observational identification of subclusters is a difficult task. The ultimate goal would be to reconstruct the merger by properly accounting for all subclusters. However, observations are all projected quantities and are limited by their sensitivity and noise. In our study, we have identified subclusters as those that have WL S/N $>3\sigma$ and coincident galaxy peaks. Although we did not include the ICM features as criteria for substructure identification, the substructures also tend to agree with the underlying ICM distribution as revealed by the XMM-\textit{Newton} observations.

\begin{figure*}
    \includegraphics[width=\textwidth]{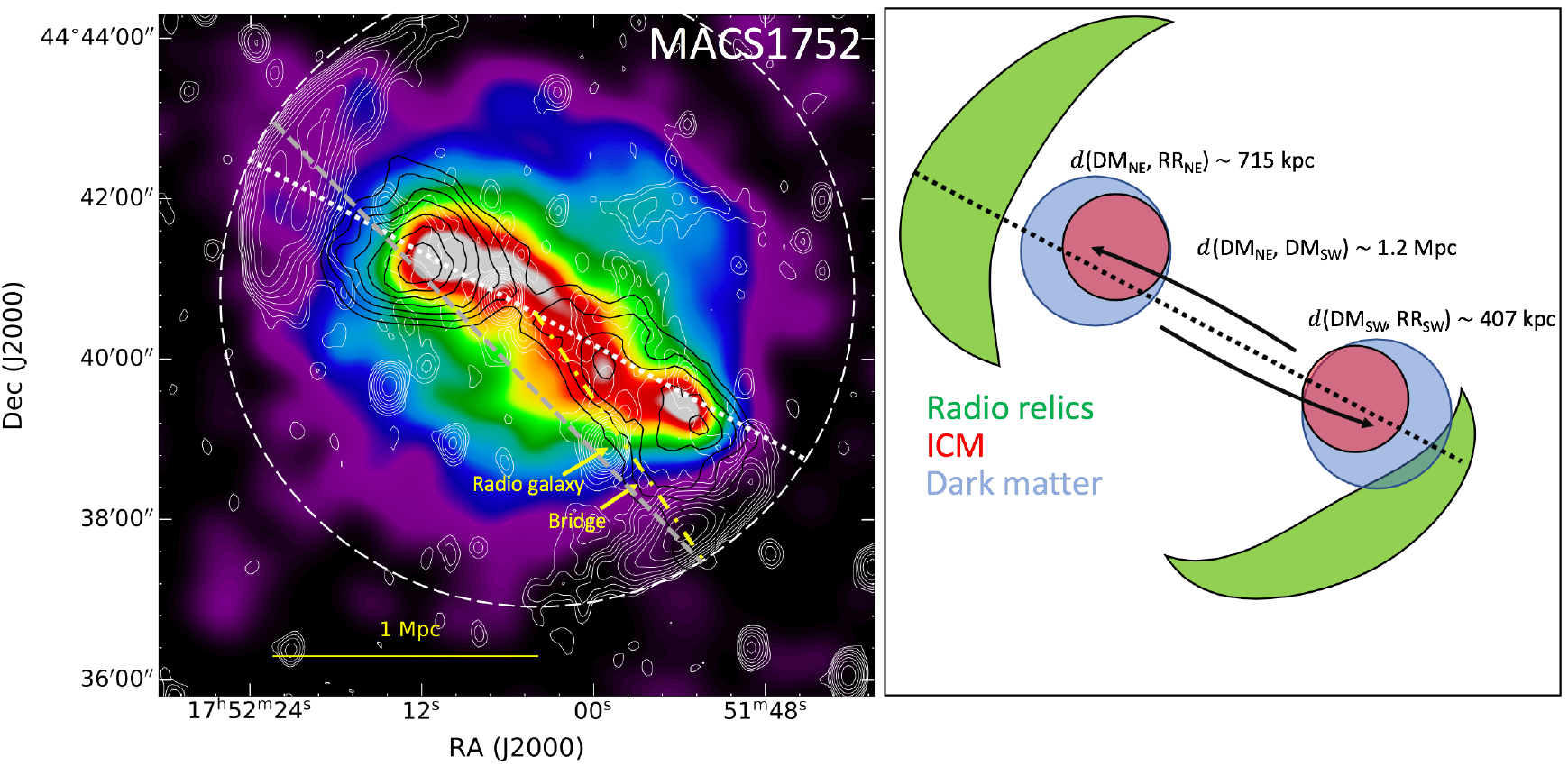}
    \caption{Left: XMM-\textit{Newton} X-ray emission (0.5-7~keV) with WSRT 18 cm radio emission contours in white and WL mass contours in black. The white dotted line is the proposed merger axis. The grey dashed line is the axis connecting the centers of the radio relics. The agreement between the mass peaks and X-ray emission peaks is clear. The NE radio relic appears to be aligned with the merger axis but the SW radio relic is pointed (yellow dash-dot branch) \mytilde25 degrees clockwise from the merger axis. Right: Schematic of the merger scenario with measured distances. A merger with a non-zero impact parameter between two equal mass subclusters is depicted.}
    \label{fig:macsj1752_merger}

\end{figure*}

The global distributions of WL mass, gas, and galaxies in MACS1752 follow a similar morphology. As revealed by the WL mass reconstruction, MACS1752 is predominantly composed of two subclusters (NE and SW) that are co-spatial with the BCGs and the X-ray brightness peaks. The elongated distribution of the bright X-ray emission from the ICM follows an inverted S shape that terminates at the subcluster centers (BCGs, mass peaks, X-ray peaks). From the NE subcluster, an ICM tail appears to have been stripped from the subcluster core (left panel of Figure \ref{fig:macsj1752_merger}). This dissociated gas extends from the NE subcluster about 0.5 Mpc towards the barycenter of the system. An X-ray tail of this extent would require a powerful ram-pressure force and is suggestive of a low impact parameter for the collision. From the SW subcluster, an extension of the gas towards the center of the cluster may also be caused by ram pressure. However, this extension also coincides with the central WL mass detection and X-ray peak. It is difficult to say whether the central substructure is involved in the merger or is a foreground substructure that will merge later. The BCG that coincides with the mass peak is at $z=0.3617$, which is slightly lower redshift than the SW ($z=0.3634$) and NE ($z=0.3648$) BCGs (Table \ref{table:mcmac}). Higher resolution or S/N X-ray observations may provide insight into the nature of this third substructure.

The WL signal of ZWCL1856 is dominated by the N and S subclusters. Based on our bootstrap analysis, the mass peaks in the N and S are consistent with the corresponding X-ray brightness peaks. The resemblance of the distributions of WL mass and X-ray emission of ZWCL1856 and MACS1752 are extraordinary. An extension of the WL mass distribution to the north of the S subcluster is apparent. This is comparable to the extension seen from the SW subcluster in MACS1752 but remains unidentified because of the lack of cluster galaxies in the region. The X-ray emission also shows evidence of gas dissociation with a connection between the subclusters that resembles the inverted S shape of MACS1752.

\begin{figure*}
    \includegraphics[width=\textwidth]{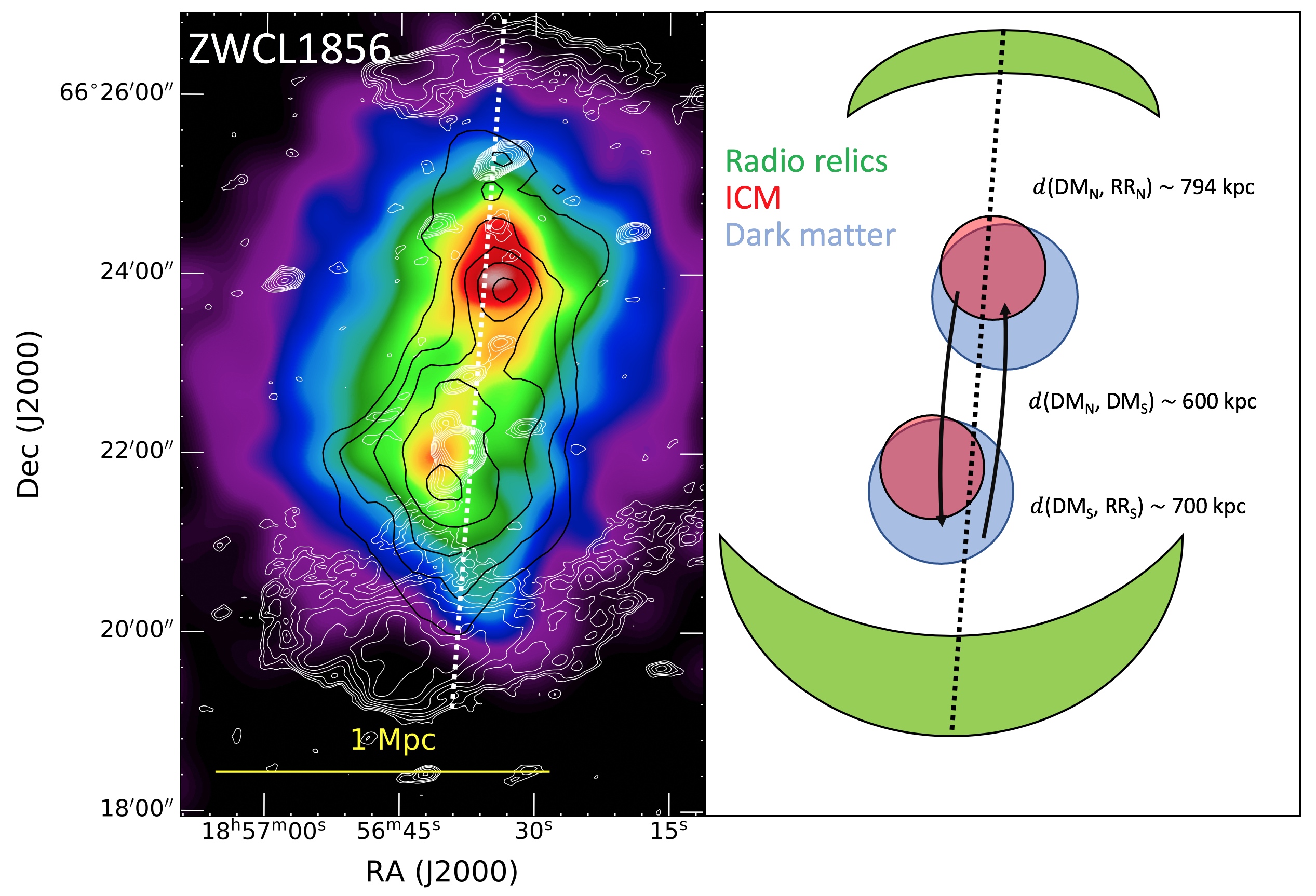}
    \caption{Left: XMM-\textit{Newton} X-ray emission (0.5-7 keV) with LOFAR 144~MHz radio emission contours in white and WL mass contours in black. The mass peak in the north lies closer to the center of the cluster than the respective X-ray brightness peak. The white dotted line is our proposed merger axis. Right: Schematic of the merger scenario with measured distances. }
    \label{fig:ZWCL1856_merger}

\end{figure*}

\subsection{Parameterizing the Merger Scenarios}
Gathering the radio information from the literature and combining it with our new WL mass maps, mass estimates, and X-ray analysis, we revisit the merger scenarios for two of the most symmetric double radio relic clusters. The merger parameters that are of particular interest to reconstructing the past merger are \edit2{the mass ratio}, the separation of the subclusters, the collision velocity and direction, the observed phase, and the time since collision (TSC, the time passed since first pericenter passage). \edit2{In \S \ref{sec:macs1752_merger} and \S \ref{sec:ZWCL1856_merger}, we discuss the merger scenarios of MACS1752 and ZWCL1856, respectively, and report estimates for the aforementioned parameters. Pros and cons of the methods used are also discussed.}

\subsubsection{MACS1752}  \label{sec:macs1752_merger}
The multiwavelength view of MACS1752, shown in the left panel of Figure \ref{fig:macsj1752_merger}, presents a binary merger between the NE and SW subclusters. Our lensing analysis found that this is a merger with 1:1 mass ratio. This mass ratio is in agreement with galaxy velocity dispersion measurements from \cite{2019bgolovich} (Table \ref{table:mass_estimations}). On the other hand, from the brightness of the radio relics, \cite{2012vanweeren} proposed a 2:1 mass ratio and in the simulation of \cite{2012bonafede} a 3:1 mass ratio was suggested. 

\edit2{The viewing angle of the merger has been discussed in previous studies.} The spectroscopic redshift study of \cite{2019bgolovich} found that a two-halo model is preferred and that the two subclusters have similar redshifts; evidence that supports MACS1752 as a plane-of-the-sky merger. \cite{2012bonafede} showed that the flattest part of the spectral index for the northeast relic is located at the outer edge of the radio emission, which is expected for a shock propagating in the plane of the sky. However, the southwest relic has a spectral index that is flattest within the radio emission, which may indicate a slight viewing angle to the radio emission \edit2{\citep{2013skillman}}. 

The path that the merger unfolded along may be inferred from the distribution of the X-ray emission \edit2{(left panel of Figure \ref{fig:macsj1752_merger})}. The schematic view of the merger is shown in the right panel of Figure \ref{fig:macsj1752_merger}. The X-ray tail that lags behind the NE subcluster is a good indicator of the path that the NE subcluster followed and suggests a collision with a non-zero impact parameter. Given the equality of the subcluster masses, we expect the SW subcluster to have followed a similarly shaped path in the opposite direction. \edit2{Evidence for the SW subcluster path may be present in the X-ray tail that extends from the SW subcluster but it blends in with the central substructure. Higher resolution X-ray imaging is needed to better define these features.}

Recreating radio relics in numerical simulations is often done by modeling the kinetic energy flux, which is proportional to the collision velocity of the subclusters. We tested three methods to approximate the collision velocity of MACS1752. \edit2{Each method also provides an estimate of the TSC.}

\begin{itemize}
\item We ran the Monte Carlo Merging Analysis Code \citep[MCMAC,][]{2013dawson} with the parameters given in Table \ref{table:mcmac}. MCMAC samples the posterior distribution of a binary, head-on collision of NFW halos with all viewing angles equally sampled. We constrained the samples to only those merging within 30 degrees of the sky \citep{2019bgolovich, 2019wittman}. \edit2{The top row of Figure \ref{fig:phase} shows the MCMAC posterior distributions of the projected shock distance (left), TSC (middle), and collision velocity (right, $V_{3D}$). Posteriors are shown as blue for the outgoing phase (between first pericenter and first apocenter) and orange for the returning phase (after first apocenter). The distance to the radio relic from the barycenter of the system is marked with a vertical black line and the uncertainty is encapsulated in the grey, dashed vertical line. Utilizing the constraining power of the observed radio relic position, we find that the system has a \mytilde24\% greater probability of being in the outgoing phase. The TSC and collision velocity of the outgoing phase are $0.9\pm0.1$ Gyrs and $2233^{+143}_{-130}$ km s$^{-1}$. } We note that the velocity (TSC) from MCMAC is expected to be biased lower (higher) than the true collision velocity (TSC) because the halos are released from their current separation and given an initial relative velocity between zero and the free-fall velocity. \cite{2019wittman} explained that the head-on collision assumption in MCMAC leads to a TSC overestimation because it forces a radial component into the merger rather than having a nonzero pericenter distance. This in turn increases the 3D separation of the subclusters and the TSC.

\item Investigating merging clusters in the BigMDPL cosmological simulations, \cite{2019wittman} found that 68\% (95\%) of MACS1752 analogs have collision velocities ranging from 2444-3034 (1737-3264) km s$^{-1}$. The analogs provide a TSC for the system of 0.2-0.5 (0.1-1.1) Gyrs. \edit2{These values are $2\sigma$ consistent with the values found from MCMAC.  The analog method also constrains the MACS1752 merger to be within 16 (30) degrees of the plane of the sky. An advantage of the analog method is that it can sample the collision velocity from the simulations, rather than infer it analytically. One of the disadvantages of the analog method is that it does not utilize the locations of the radio relics.}

\item The radio relics can be used to approximate the collision speed by assuming the DSA model and converting the spectral indices to Mach numbers. \cite{2018ha} showed that axial shocks propagate at a nearly constant speed and Mach number from impact to 1 Mpc, which suggests that the shock Mach number may be a valid proxy for the collision speed. For MACS1752, \cite{2012bonafede} reported injection spectral indices of 0.6 and 0.8, which give Mach numbers of 4.6 and 2.8 for the NE and SW radio relics, respectively. Unable to measure the X-ray temperature in the pre-shock region (outside the radio relic), we calculated the ICM sound speed from the global temperature (\mytilde$5.9$ keV) to be $c_s\sim1200$ km s$^{-1}$. This should be taken as an upper limit on the sound speed assuming that the merger activity has inflated the temperature of the cluster. Based on these properties, the expected collision velocity is 5520 km s$^{-1}$ for the NE relic and 3360 km s$^{-1}$ for the SW. These high velocities translate to a TSC in the range 0.21 - 0.35 Gyrs for the average relic distance of 1.2 Mpc from barycenter. \edit2{It is worth noting that the collision velocity and TSC from the radio relics are somewhat departed from the MCMAC and analog methods.} A number of systematic effects may play a role \citep{2014stroe, 2016vanweeren, 2017hoang}. Furthermore, Mach numbers inferred from radio spectral indices tend to be high, which may be caused by a preferential detection of the highest Mach number region of the shock \citep{2018ha, 2019vanweeren}.   

\end{itemize}


\begin{table}[]
\caption {MCMAC input parameters} \label{table:mcmac}
\def\arraystretch{1.}
\begin{tabular}{ c c c c }
\hline
\hline
Parameter  & Value & Uncertainty & Units          \\
\hline
\textbf{MACS1752}    &         &                       \\
$M_{\mathrm{NE}}$  & 5.6     & 1.7   & $10^{14}\ $M$_\odot$          \\
$M_{\mathrm{SW}}$ & 5.6     & 1.8  & $10^{14}\ $M$_\odot$            \\
$d_{\mathrm{sep}}$  & 1.2 & 0.1   & Mpc            \\
$z_{\mathrm{NE}}$      & 0.3648 & 0.0005 &  \\
$z_{\mathrm{SW}}$     & 0.3634 & 0.0005 & \\
\hline
\textbf{ZWCL1856} &         &                       \\
$M_{\mathrm{N}}$  & 1.2     & 0.5       & $10^{14}\ $M$_\odot$      \\
$M_{\mathrm{S}}$  & 1.0     & 0.5       & $10^{14}\ $M$_\odot$      \\
$d_{sep}$    & 0.6 & 0.1   & Mpc          \\
$z_\mathrm{N}$      & 0.3041 & 0.0009 & \\
$z_\mathrm{S}$      & 0.3033 & 0.0007 & \\
\hline

\end{tabular}

\end{table}

\subsubsection{ZWCL1856} \label{sec:ZWCL1856_merger}
ZWCL1856 (Figure \ref{fig:ZWCL1856_merger}) has a similar ICM morphology to MACS1752 with the X-ray emission showing a bright, inverted S-shaped region that runs between the subclusters \edit2{(left panel of Figure \ref{fig:ZWCL1856_merger})}. The ICM morphology does not show the prominent gas tails that are found in MACS1752 but the bright emission located between the subclusters could be from gas dissociation. As in MACS1752, the S-shaped feature indicates a non-zero impact parameter for the collision. The radio relics have a strong symmetry with the elongation of the ICM. Furthermore, \cite{2014degasperin} suggested that the viewing angle of ZWCL1856 may be slightly tilted from the plane of the sky because of the detection of strong polarization in the northern relic and little polarization in the southern.

 Our WL result portrays the merger as about 1:1 mass ratio but at a much lower total mass than MACS1752. It is expected that the lower mass of the system should decrease the collision velocity. We use the same three methods as above to constrain the collision velocity \edit2{and TSC} of ZWCL1856. 
\begin{itemize}

\item \edit2{The MCMAC result for ZWCL1856 is presented in the bottom row of Figure \ref{fig:phase}. The left panel shows that the returning phase is $99.5\%$ more likely than the outgoing phase when constricted to be within the uncertainty of the two radio relic positions. For the returning phase, the TSC is $1.7^{+0.2}_{-0.1}$ Gyrs and the collision velocity is $1367^{+92}_{-101}$ km s$^{-1}$. We reiterate that MCMAC tends to overestimate TSC and underestimate collision velocity.}  

\item The 68\% (95\%) distribution of simulated analogs of \cite{2019wittman} prefer a collision velocity of 1494-1866 (1330-2094) km s$^{-1}$ and a TSC of 0.3-0.5 (0.1-0.6) Gyrs. \edit2{Although the collision velocity is consistent with the MCMAC result, the TSC is inconsistent. The analog method returned a majority of mergers that were in the outgoing phase. Again, including the radio relic locations in the analog method would improve the selection of analogs.} The viewing angle of ZWCL1856 is constrained to be within 24 (44) degrees of the plane of the sky from analogs.  

\item \edit2{From LOFAR, VLA, and GMRT radio observations, Jones et al. submitted. measured the spectral index along the shock front to derive Mach numbers of $M=2.5\pm0.2$ and $M=2.3\pm0.2$ for the north and south shocks, respectively. Given a 3.7 keV plasma with a sound speed of \mytilde 1000 km s$^{-1}$, the collision velocity is 2100 to 2700 km s$^{-1}$. These give a range of TSC from 0.40-0.51 Gyrs for the relics at 1.1 Mpc from barycenter.} 

\end{itemize}

MCMAC, simulated analogs, and the radio relics portray MACS1752 as a higher-speed collision than ZWCL1856. This is in agreement with our expectation from the mass estimates. \edit2{When it comes to the TSC, all three methods predict that ZWCl1856 is an older merger than MACS1752. However, this is only true for the central value of the analog method as the uncertainties overlap.}

\subsubsection{Comparing the Merger Phases of MACS1752 and ZWCL1856} \label{sec:comparison}
The two clusters that we have analyzed in this work have remarkably similar features, \edit2{even in comparison to other diametric radio relic clusters \citep[see Figure 16 of][]{2019vanweeren}}. Both are post-merger systems with double radio relics and bimodal mass distributions. Both systems have evidence of ICM interactions from a collision with a small but non-zero pericenter distance. However, an analysis of the differences will shed light on the merger phases.

\begin{figure*}
    \includegraphics[width=\textwidth]{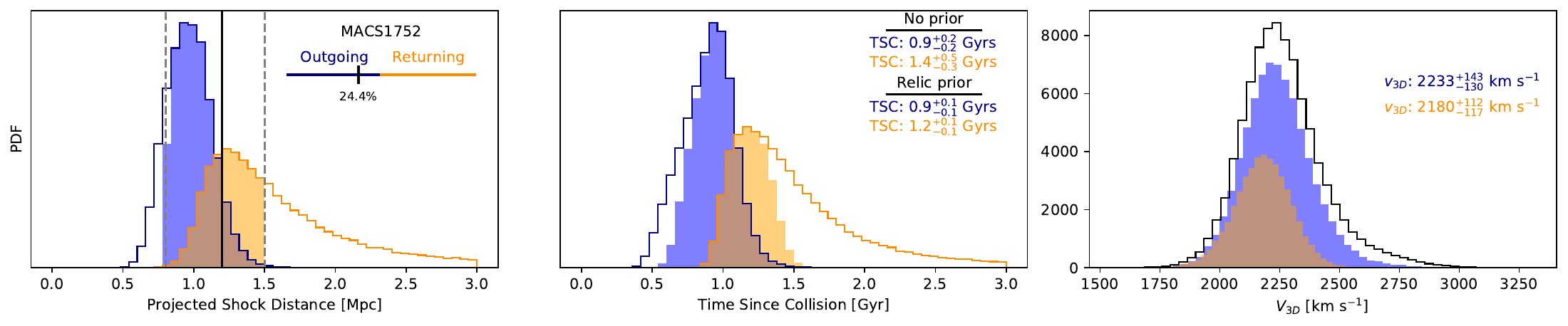}\\
    \includegraphics[width=\textwidth]{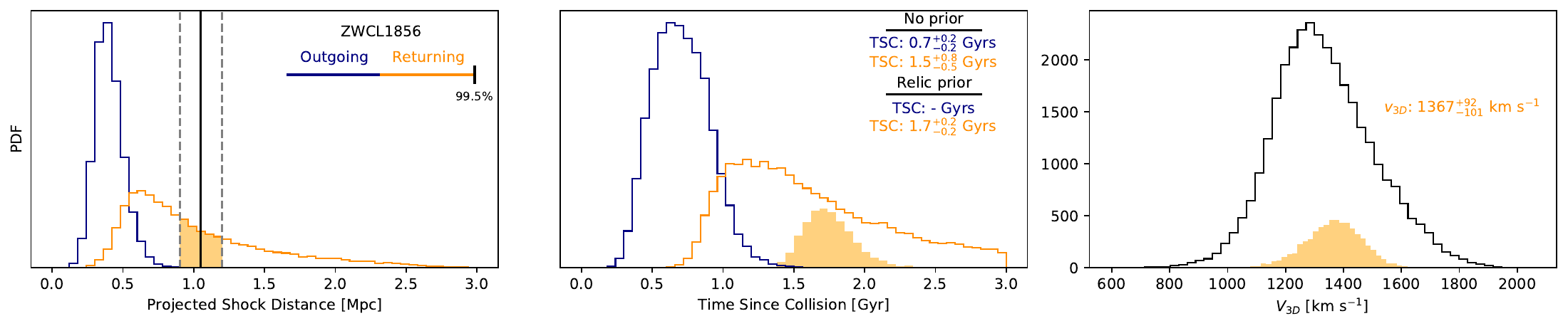}
    \caption{Phase predictions (left), TSC (middle), and collision velocity (right) for the MACS1752 (top) and ZWCL1856 (bottom) mergers from the posterior distributions of MCMAC. The black line represents the observed position of the relics with respect to the barycenter of the cluster. \edit2{The relic position $1\sigma$ uncertainties are shown as grey, dashed lines. The shaded PDFs represent the realizations that are within the prior of the relic position. Probability of merger being in the outgoing and returning phases are visualized with the meter. }}
    \label{fig:phase}
\end{figure*}

\edit2{Figure \ref{fig:phase} presents the expected phase of the merger based on the MCMAC posterior distributions and the measured positions of the radio relics. The posterior distributions indicate that MACS1752 is between first pericenter and first apocenter. On the other hand, ZWCL1856 is further progressed in its merger and observed returning from first apocenter. In the remainder of this section, we provide additional indicators that may shed light on the merger phase. }

An indication that ZWCL1856 is at a later stage of merging than MACS1752 comes from the comparison of the geometry of the systems. We summarize the projected distances between the observed features of the clusters in Table \ref{table:distances}. These measurements were made parallel to the defined collision axis (the white dotted lines in Figures \ref{fig:macsj1752_merger} and \ref{fig:ZWCL1856_merger}). We define the ratio of the radio relics separation to the subcluster WL peaks separation as R2D (Table \ref{table:distances}). R2D is a ratio that should almost always be increasing with time between first and second pericenter passage. Only in mergers that are observed near the radial direction from the observer and with a large impact parameter (ie. a large turn in the merger path) could the subclusters appear to outrun the radio relics and in that case it is expected that the radio relics would not be visible. The values of R2D that we have found for MACS1752 and ZWCL1856 are 1.9 and 3.5, respectively. \edit2{R2D will also vary with other cluster properties such as the gas density and dark matter concentration. However, the larger R2D of ZWCL1856 is in agreement with previous evidence that we have presented that shows ZWCL1856 is in a later stage of merging than MACS1752.}

Another indication of the phase of the merger is the standoff distance \citep{2003verigin, 2019zhang}, which compares the radius of a cold front to the separation distance of a shock from the stagnation point (leading edge of cold front / contact discontinuity). Unable to detect a cold front in the X-ray emission of ZWCL1856, we instead use the distance from the X-ray brightness peak to the radio relic as a proxy for standoff distance. The X-ray peak to relic distances of the ZWCL1856 subclusters are \mytilde700 kpc and \mytilde794 kpc. The NE relic of MACS1752 is \mytilde715 kpc from its X-ray brightness peak and the SW relic is \mytilde407 kpc. Given the same time interval, one would expect the radio relics of MACS1752 to separate more because of the higher impact velocity (shock speed). \edit2{Since the total masses of ZWCL1856 and MACS1752 are quite different, comparing the WL peaks to radio relic separations alone is not a strong indication of the merger state, but we will note that ZWCL1856 has a larger average X-ray peak to relic distance than MACS1752.}

\begin{table}[]
\caption {Projected Separations and Merger Properties} \label{table:distances}
\def\arraystretch{1.}
\begin{tabular}{ c c c c }
\hline
\hline
Parameter & Value & Units            \\
\hline
\textbf{MACS1752}    &         &                       \\
$d($DM$_{\mathrm{NE}}$, DM$_{\mathrm{SW}})$ & 1200 & kpc \\
$d($DM$_{\mathrm{NE}}$, RR$_{\mathrm{NE}})$ & 715 & kpc \\
$d($DM$_{\mathrm{SW}}$, RR$_{\mathrm{SW}})$ & 407 & kpc \\
$d($RR$_{\mathrm{SW}}$, RR$_{\mathrm{NE}})$ & 2323 & kpc \\
$v_c$ & 2444-3034 & km s$^{-1}$ \\
TSC & 0.2-0.5 & Gyrs \\
R2D & 1.9 & \\
\hline
\textbf{ZWCL1856} &         &                       \\
$d($DM$_\mathrm{N}$, DM$_\mathrm{S})$ & 600 & kpc \\
$d($DM$_\mathrm{N}$, RR$_\mathrm{N})$ & 794 & kpc \\
$d($DM$_\mathrm{S}$, RR$_\mathrm{S})$ & 700 & kpc \\
$d($RR$_\mathrm{S}$, RR$_\mathrm{N})$ & 2094 & kpc \\
$v_c$ & 1494-1866 & km s$^{-1}$ \\
TSC & 0.3-0.5 & Gyrs \\
R2D & 3.5 & \\
\hline

\end{tabular} \\
DM = dark matter peak (weak lensing) \\
RR = center of radio relic shock front \\
$v_c$ and TSC are adopted from \cite{2019wittman} analogs \\
R2D is defined in Section \ref{sec:comparison}
\end{table}

\subsection{A Case for Non-thermal Particle Re-acceleration} \label{sec:reacceleration}
The rarity of radio relics in merging galaxy clusters and the inefficiency of DSA has led to the re-acceleration theory gaining popularity. Evidence for re-acceleration has been provided in work such as \cite{2012bonafede} and \cite{2017vanweeren}. In \cite{2017vanweeren}, they show the clearest evidence for re-acceleration with a bridge connecting a nearby AGN to the radio relic. We propose that re-acceleration may be observed in MACS1752.

We simplify the merger scenario by defining a collision axis. With the absence of a consensus on the definition of a collision axis from literature, we propose that the collision axis should be defined from the antiparallel velocity vectors that connect the dark matter halos at pericenter. However, one cannot directly observe the subcluster motions and must find a collision-axis proxy. One method is to utilize the positions and orientations of the radio relics as a proxy. In some double radio relic clusters (e.g. CIZA J2242.8+5301, ZWCL 0008.8+5215), connecting the centers of the radio relics with a line gives an approximation of the collision axis that is comparable to the expected axis at pericenter. This is also demonstrated for shocks in the simulations of \cite{2001ricker} and \cite{2011zuhone} but may not necessarily be true for radio relics if they depend on the local ICM properties. For MACS1752, reconciling the collision axis defined by the radio relics (grey dotted line in left panel of Figure \ref{fig:macsj1752_merger}) with the X-ray morphology and the observed locations of the WL peaks is difficult. Instead, we suggest that the collision occurred offset from the axis defined by the radio relics. As an alternative, we take the major axis of an elliptical Gaussian fit to the X-ray emission as the collision axis. This new collision axis (heavier white dotted line in Figure \ref{fig:macsj1752_merger}) bisects the WL and X-ray peaks and is \mytilde25 degrees rotated counter-clockwise from the relic-defined axis. For robustness, we also tested elliptical Gaussian fits to the galaxy and WL distributions and found them to indicate a collision axis that is similar to that given by the X-ray distribution. 

Assuming that the relic should be visible off the bow of the subcluster, a slight tension with the position of the SW radio relic is found. We theorize that this may be a case for re-acceleration of a non-thermal population of cosmic rays. Figure 5 in \cite{2012bonafede} shows a connection between the cluster radio galaxy and radio relic in 323~MHz GMRT observations at the $3\sigma$ level. This bridge is also shown in the WSRT 18 cm contours in our Figure \ref{fig:macsj1752_merger}. A line from the barycenter of the merger to the center of the radio relic passes directly through this radio galaxy. We confirm that this galaxy is a cluster member at a redshift of $0.37086\pm0.00002$ from the Keck observations of \cite{2019bgolovich}. In our merger scenario, the shock is launched along the collision axis in the direction of motion of the subcluster. We theorize that the south-west facing portion of the shock passed through the radio galaxy and has been re-accelerating the non-thermal electron population. \cite{2012bonafede} presented a spectral index map of MACS1752 that shows a hint of a gradient that connects the radio galaxy to the radio relic but the evidence is not strong. New radio observations may provide the resolution to clearly see the spectral signature of a connection between the galaxy and the relic. In addition, deeper X-ray observations to search for the shock at the position of the radio relic and \edit2{directly west} of the SW subcluster would be revealing.

\subsection{\textit{HST} Strong-lensing Features}
\begin{figure}[t]
    \centering
    \includegraphics[width=0.45\textwidth]{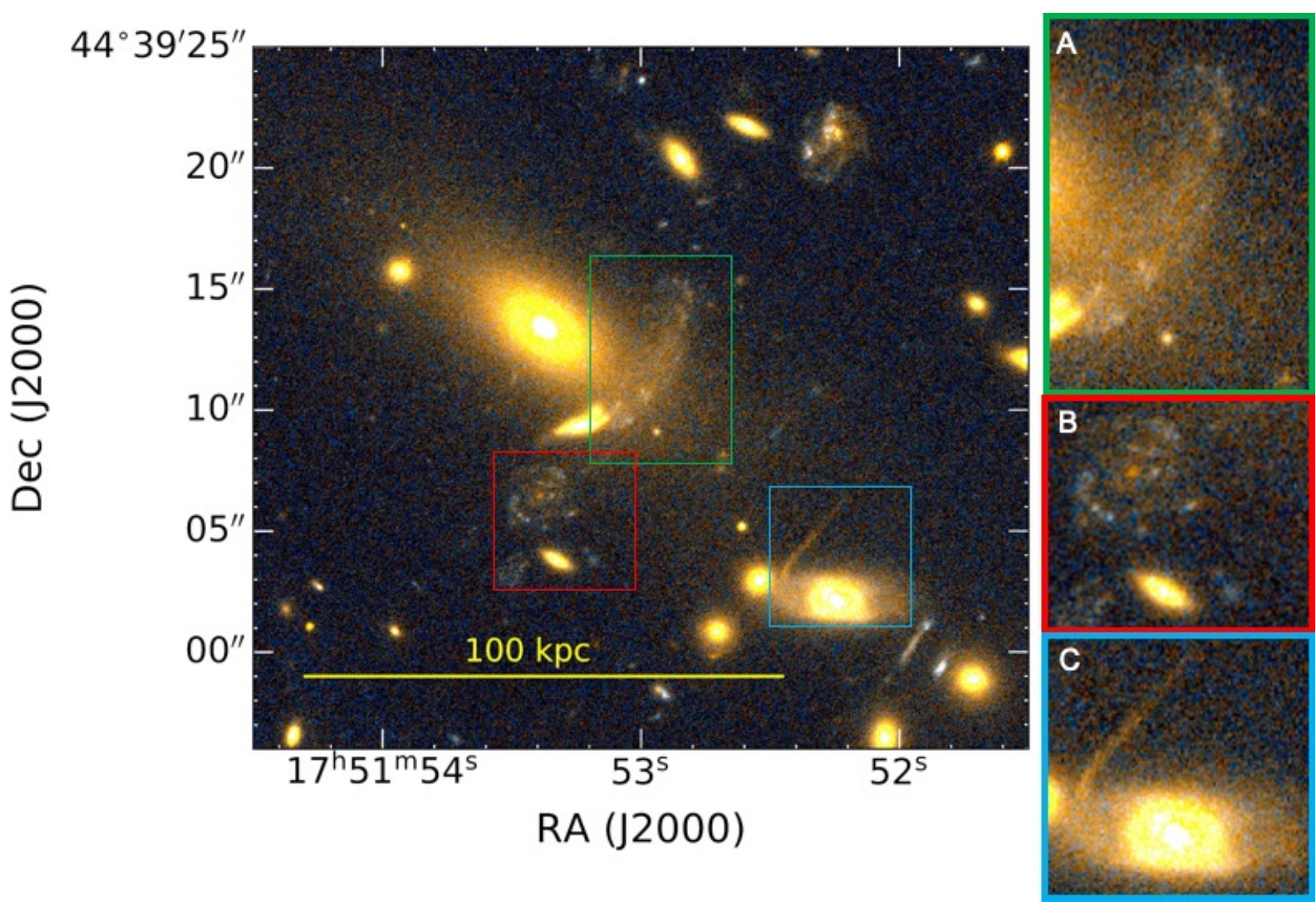}
    \caption{Strong-lensing image candidates near the SW BCG of MACS1752. (A) A face-on spiral galaxy image. (B) A galaxy that exhibits similar features and color to image A. (C) A strong-lensing arc that curves away from the BCG.}
    \label{fig:strong_lens}
\end{figure}

The deep, high-resolution \textit{HST} imaging provides an opportunity to find strong-lensing features in the vicinity of the MACS1752 subclusters. Figure \ref{fig:strong_lens} shows the region around the SW BCG that contains candidate strong-lensing images. Overlapping the light of the BCG, there is a lensed, face-on spiral galaxy (green box). This light is partially blocked by the galaxy companion to the BCG. A similar face-on spiral galaxy (red box) is located southeast of the lensed galaxy but it does not have the tell-tale features of a lensed galaxy. Located further south east is a strong-lensing arc (blue box) that bends away from the BCG. The bend may be caused by the prominent cluster galaxy that is shown within the blue box. Future studies could model these features to constrain the small-scale mass distribution.

\section{Conclusions} \label{sec:conclusion}
We have presented a weak-lensing and X-ray analysis of two merging galaxy clusters that exhibit double radio relics, MACSJ1752.0+4440 and ZWCL1856.8+6616. Our analysis quantitatively constrained the merger scenarios of these two remarkably similar merging galaxy clusters. The constraints summarized below will be valuable for future simulations of cluster mergers.    

\subsection*{\textbf{MACSJ1752.0+4440}}
\begin{itemize}
    \item Our weak-lensing analysis of Subaru and \textit{HST} observations have revealed a double-peaked mass distribution. These mass peaks are consistent with the X-ray brightness peaks and galaxy peaks.
    
    \item Modeling each subcluster as an NFW halo, we have concluded that MACS1752 has a mass ratio of 1:1 with the NE (SW) subcluster mass being $M_{200} = 5.6^{+1.8}_{-1.6}\times 10^{14}\ $M$_\odot$ ($M_{200} = 5.6^{+1.4}_{-2.1}\times 10^{14}\ $M$_\odot$). The total weak-lensing mass is $M_{200}=14.7^{+3.8}_{-3.3}\times10^{14}\ $M$_\odot$ and is consistent with the Planck SZE mass.
    
    \item XMM-\textit{Newton} observations show gas tails and an inverted S-shape that connects the X-ray brightness peaks. These features are interpreted as past signs of merging and indicators of a nearly head-on collision. 
    
    \item \edit2{Using the posterior distributions of MCMAC and the locations of the radio relics, we have found that MACS1752 is in the outgoing phase, between first pericenter and apocenter.}
    
    \item The radio relics, subcluster masses derived from weak lensing, and geometry of the system are used to investigate the merger scenario. \edit2{Constraints on the phase, TSC, and collision velocity are explored with MCMAC, analogs from the BigMDPL, and the Mach numbers and positions of the radio relics. Our analysis predicts that MACS1752 is in the outgoing phase with TSC estimates of $0.9\pm0.2$ Gyrs, $0.2-0.5$ Gyrs, and $0.21-0.35$ Gyrs for the MCMAC, analogs, and radio relics methods, respectively. These methods find $2233^{+143}_{-130}$ km s$^{-1}$, $2444-3034$ km s$^{-1}$, and $3360-5520$ km s$^{-1}$ for the velocity of the collision.  The pros and cons of each method are discussed.}
    
    \item Defining the merger axis from the X-ray distribution, we have postulated that the SW radio relic is rotated from its expected position and may be a signature of particle re-acceleration in action. Further radio and X-ray observations are required to make stronger conclusions.  
\end{itemize}

\subsection*{\textbf{ZWCL1856.8+6616}}
\begin{itemize}
    \item Our weak-lensing analysis of Subaru imaging has found that ZWCL1856 is also a double-peaked merging cluster with a 1:1 mass ratio but at a much lower total mass than MACS1752. 
    
    \item From simultaneously fitting two NFW halos to the lensing signal, we estimate the mass of ZWCL1856 N (S) to be $M_{200} = 1.2\pm0.5\times 10^{14}\ $M$_\odot$ ($M_{200} = 1.0^{+0.4}_{-0.7}\times 10^{14}\ $M$_\odot$) and total mass to be $M_{200}=2.4^{+0.9}_{-0.7}\times10^{14}\ $M$_\odot$. This mass estimate is lower than the SZE mass estimate from Planck. However, merging systems like ZWCL1856 consist of two subclusters that have recently undergone an energetic collision and one should expect scatter in the relation of weak-lensing mass and SZE mass. \\
    
    \item XMM-\textit{Newton} observations of ZWCL1856 are presented. Qualitatively, the X-ray emission and weak-lensing distributions of ZWCL1856 look similar to MACS1752. The bright X-ray emission has a similar inverted S shape to that found in MACS1752. In contrast to MACS1752, the radio relics of ZWCL1856 are well aligned with the elongated X-ray morphology and weak-lensing mass distribution. 
    
    \item \edit2{MCMAC predicts that ZWCL1856 is in the returning from apocenter phase of the merger. }
    
    \item \edit2{Our tests show that the ZWCL1856 merger was a slower collision than MACS1752 with collision velocity estimations of $1367^{+92}_{-101}$ km s$^{-1}$, $1494-1855$ km s$^{-1}$, and $2100-2700$ km s$^{-1}$ from MCMAC, analogs, and radio relics, respectively. The TSC estimations from the three methods are $1.7^{+0.2}_{0.1}$ Gyrs, $0.3-0.5$ Gyrs, and $0.40-0.51$ Gyrs. } 
    
\end{itemize}

We thank the referees for carefully reading the manuscript and providing comments to improve it. This  work  is  based  on  observations  made  with  the NASA/ESA  {\it Hubble  Space  Telescope}  and  operated  by the Association of Universities for Research in Astronomy,  Inc.   under  NASA  contract  NAS  5-2655.
MJJ acknowledges support from the National Research Foundation of Korea  under  the  program nos. 2017R1A2B2004644 and 2017R1A4A1015178. 
Part of this work was performed under the auspices of the U.S. Department of Energy by Lawrence Livermore National Laboratory under Contract DE-AC52-07NA27344.

\software{SDFRED2 \citep{2002yagi, 2004ouchi}, SExtractor \citep{1996bertin}, SCAMP \citep{bertinscamp}, SWARP \citep{bertinswarp}, Multidrizzle \citep{2003koekemoer}, XMM-SAS \citep[v18.0.0;][]{2004gabriel}, FIATMAP \citep{1997fischer}, MCMAC \citep{2013dawson}}
\bibliography{mybib.bib}{}
\bibliographystyle{aasjournal.bst}

\end{document}